Volume conservation during finite plastic deformation

He-Ling Wang[1, 2], Dong-Jie Jiang[1], Li-Yuan Zhang[1], Bin Liu[1] *

[1]*AML, CNMM, Department of Engineering Mechanics, Tsinghua University, Beijing, 100084, China*
[2] *Department of Civil and Environmental Engineering, Northwestern University, Evanston, IL 60208, USA.*

* Corresponding author, Tel.: 86-10-62786194; fax: 86-10-62781824
E-mail address: liubin@tsinghua.edu.cn (B. Liu);

Abstract

An elastoplastic theory is not volume conserved if it improperly sets an arbitrary plastic strain rate tensor to be deviatoric. This paper discusses how to rigorously realize volume conservation in finite strain regime, especially when the unloading stress free configuration is not adopted or unique in the elastoplastic theories. An accurate condition of volume conservation is clarified and used in this paper that the density of a volume element after the applied loads are completely removed should be identical to that of the initial stress free states. For the elastoplastic theories that adopt the unloading stress free configuration (i.e. the intermediate configuration), the accurate condition of volume conservation is satisfied only if specific definitions of the plastic strain rate are used among many other different definitions. For the elastoplastic theories that do not adopt the unloading stress free configuration, it is even more difficult to realize volume conservation as the information of the stress free state lacks. To find a universal approach of realizing volume conservation for elastoplastic theories whether or not adopt the unloading stress free configuration, we propose a single assumption that the density of material only depends on the trace of the Cauchy stress, and interestingly find that the zero trace of the plastic stress rate is equivalent to the accurate condition of volume conservation. Two strategies are further proposed to satisfy the accurate condition of volume conservation: directly and slightly revising the tangential stiffness tensor or using a properly chosen stress/strain measure and elastic compliance tensor. They are implemented into existing elastoplastic theories, and the volume conservation is demonstrated by both theoretical proof and numerical examples. The potential application of the proposed theories is a better simulation of manufacture process such as metal forming.

Table of nomenclature

| | |
|---|---|
| $\boldsymbol{F}$ | Deformation gradient |
| $\boldsymbol{F}^{\mathrm{e}}$ | Elastic deformation gradient: deformation gradient from the unloading stress free configuration to the current configuration |
| $\boldsymbol{F}^{\mathrm{p}}$ | Plastic deformation gradient: deformation gradient from the initial stress free configuration to the unloading stress free configuration |
| $\boldsymbol{C}^{\mathrm{p}}$ | Plastic right Cauchy–Green tensor |
| $\boldsymbol{E}$ | Green strain |
| $\boldsymbol{E}^{\ln}$ | Logarithmic strain |
| $\boldsymbol{E}^{(n)}$ | Seth's strain |
| $\varepsilon$ | Strain of small deformation |
| $\bar{\boldsymbol{E}}$ | Strain by taking the unloading stress free configuration as the reference configuration |
| $\dot{\boldsymbol{E}}$ | Strain rate |
| $\dot{\bar{\boldsymbol{E}}}$ | Strain rate by taking the unloading stress free configuration as the reference configuration |
| $\boldsymbol{d}$ | Deformation rate |
| $\left(\dot{\boldsymbol{E}}^{\mathrm{SO}}\right)_{\mathrm{p}}$ | Plastic strain rate suggested by the Simo–Ortiz theory |
| $\left(\dot{\boldsymbol{E}}^{\mathrm{RH}}\right)_{\mathrm{p}}$ | Plastic strain rate suggested by the Rice–Hill theory |
| $\left(\dot{\boldsymbol{E}}^{\mathrm{MOS}}\right)_{\mathrm{p}}$ | Plastic strain rate suggested by the Moran–Ortiz–Shih theory |
| $\left(\dot{\boldsymbol{E}}^{\mathrm{SO}}\right)_{\mathrm{e}}$ | Elastic strain rate suggested by the Simo–Ortiz theory |
| $\left(\dot{\boldsymbol{E}}^{\mathrm{RH}}\right)_{\mathrm{e}}$ | Elastic strain rate suggested by the Rice–Hill theory |
| $\left(\dot{\boldsymbol{E}}^{\mathrm{MOS}}\right)_{\mathrm{p}}$ | Elastic strain rate suggested by the Moran–Ortiz–Shih theory |
| $\boldsymbol{\sigma}^{\mathrm{nominal}}$ | Nominal stress |
| $\boldsymbol{\sigma}^{\mathrm{Cauchy}}$, $\boldsymbol{\sigma}$ | Cauchy stress (true stress) |
| $\boldsymbol{\tau}$ | Kirchhoff stress |
| $\boldsymbol{\sigma}^{\ln}$ | Seth's stress with $n=0$ (work conjugate stress to the logarithmic strain) |
| $\boldsymbol{\sigma}^{(n)}$ | Seth's stress |
| $\left(\dot{\boldsymbol{\sigma}}\right)_{\mathrm{p}}$ | Plastic stress rate |
| $\rho$ | density |
| $A$ | Area of the cross section |
| $V$ | volume |
| $\lambda_i\left(i=1,2,3\right)$ | Stretch ratio |
| $J$ | Volume ratio |
| $\dot{\rho}$ | The derivative of a variable $\rho$ with respect to time |
| $\boldsymbol{E}_{(\cdot)}, \rho_{(\cdot)}$ | Quantities in a certain configuration |
| 0, ini | Initial stress free configuration |
| cur | Current configuration |
| sf | Unloading stress free configuration (intermediate configuration) |

1. Introduction

In elastoplastic constitutive theories, one of the most fundamental and important basis is that the plastic deformation does not change the volume of material. However, in the regime of finite deformation, this volume conservation is not rigorously realized by many theories especially those that do not adopt the multiplicative decomposition of the deformation gradient and the unloading stress free configurations, although various elastoplastic theories have been developed from different standpoints. First of all, the definition of the volume change due to plastic deformation (or the plastic volume deformation) should be expressed clearly and easy to check. In the infinitesimal deformation regime, the plastic volume deformation can be unambiguously defined as the trace of the plastic strain tensor, since all strain definitions converge. But in the finite deformation regime, because the strain and the plastic strain (or their rates) may have different definitions in different elastoplastic theories, it can be expected that the plastic volume deformation has many different definitions correspondingly if it is still directly defined by the plastic strain. Some literatures regarding to the elastoplastic theories of finite deformation are listed here for readers' reference (Mandel 1971, 1973, 1974; Rice 1971, 1975; Hill and Rice 1972, 1973; Moran et al. 1990; Besseling and Van der Giessen 1993; Yang et al. 2006; Lele and Anand 2009; Rubin and Ichihara 2010; Vladimirov et al. 2010; Volokh 2013; Altenbach and Eremeyev 2014; Shutov and Ihlemann 2014). There are also plenty of books introducing elastoplastic theories (e.g., Bertram 2005, Dunne and Petrinic 2005, Hashiguchi 2009, Lubarda 2010), and review articles presenting a comprehensive classification and discussion of representative theories (Naghdi 1990, Xiao et al. 2006). The conventional elastoplastic theories are usually based on the definition of plastic strain rate, the hardening function and the flow rule. Recently there were also models that did not need a definition of plastic strain rate and implicitly expressed the constitutive relationship in terms of stresses and strains (Rajagopal and Srinivasa 2015). Elastoplastic constitutive models were also developed for composite materials (Hong 2014; Balieu and Kringos 2015) and some advanced materials, such as ferroelectric ceramics (Chen et al. 2013a–c, 2015) shape memory alloys (Ziolkowski 2007, Thamburaja 2010, Arghavani et al. 2011), magnetic shape memory alloys (LaMaster et al. 2014) and proteins in biomaterials (Tang et al., 2007).

In the elastoplastic theories that adopt the multiplicative decomposition and the unloading stress free configuration, the volume conservation can be realized by carefully defining the plastic strain rate and set it to be deviatoric. The volume conservation condition during plastic deformation can be stated as $\det\left(\boldsymbol{C}^{\mathrm{p}}\right)=1$ where $\boldsymbol{C}^{\mathrm{p}}$ is the plastic right Cauchy–Green tensor (Miehe et al., 2002; Vladimirov et al., 2010). In rate form, the condition is $\mathrm{tr}\left(\dot{\boldsymbol{F}}^{\mathrm{p}}\cdot\left(\boldsymbol{F}^{\mathrm{p}}\right)^{-1}\right)=0$, where $\boldsymbol{F}^{\mathrm{p}}$ is the plastic deformation gradient. Accordingly, volume conserved elastoplastic theories are established and in numerical sense the exponential map algorithm is developed to properly preserve plastic incompressibility (Weber and Anand, 1990; Simo, 1992; Reese and Govindjee, 1998; Dettmer and Reese, 2004; Reese and Christ, 2008; Vladimirov et al., 2008). However whether it is proper to adopt the unloading stress free configuration in elastoplastic theories is still in question, because this configuration sometimes is not unique for different unloading paths, unreachable when the plastic deformation occurs during unloading, and causes incompatibility when the deformation is not homogeneous (Xiao et al., 2006). Therefore the question remains that how to rigorously and properly realize plastic volume conservation with extended applicability to theories whether or not adopt the unloading stress free configuration.

This paper is aimed at answering the above question and proposing strategies that can rigorously realize volume conservation without using the unloading stress free configuration. In Section 2, we first clarify an accurate condition of volume conservation that is clear and unambiguous as the benchmark throughout this paper, followed by theoretical and numerical evaluations of some classical elastoplastic theories and software according to this condition in Section 3. We find that among the evaluated theories, the tradition way of setting the plastic strain rate to be a deviatoric tensor is only valid in two theories that utilize the unloading stress free configuration and have strong assumptions on the material behavior. Then in Section 4, after abandoning the unloading stress free configuration we propose two new strategies of realizing volume conservation, with numerical implements and demonstrating examples. The strategies are compared and discussed in Section 5. Finally Section 6 concludes the paper by summarizing the major points.

## 2. The accurate volume conservation condition

An elastoplastic theory is volume conserved if it predicts no volume change when the applied loads are removed. There are two possible complexities during the unloading process of a structure or a solid: residual stress due to non-uniform deformation and non-unique unloading configurations arising from different unloading plastic deformation paths. Therefore, to clarify the meaning of volume conservation in the regime of finite deformation, we first distinguish three levels of volume conservation condition here.

Level 1: For a volume element subject to uniform deformation, its density change between the UNIQUE unloading stress free configuration (without any reverse plastic deformation) and the initial stress free configuration is zero. At this level, it is assumed that for a loaded current configuration, the corresponding unloading stress free configuration is unique and serves as a base to realize volume conservation.

Level 2: Realizing that sometimes different unloading paths and the reverse plastic deformation can lead to many different unloading stress free configurations corresponding to the same current loaded configuration, a more strict condition should be stated that for a volume element subject to uniform deformation, all of its unloading stress free configurations should have the same density as the initial stress free configuration.

Level 3: An even more strict volume conservation condition can be stated that for a structure under arbitrary deformation, after the applied loads are totally removed through any unloading path, the volume (or overall average density) of the structure should be the same as the volume (or density) before any load has been applied. In this condition, we do not rule out the residual stresses that can arise from the non-uniform deformation after unloading.

The level 1 condition has been realized and satisfied by some elastoplastic theories that adopt the unloading stress free configuration. Usually these theories assume that the unloading stress free configuration is unique for a current configuration, which is sometimes too strong. The level 2 condition has much wider applicability without the strong assumptions on the unloading process. The level 1 condition is automatically satisfied if the level 2 condition is satisfied. The level 3 condition can only be satisfied by satisfying level 2 condition and assuming a linear relationship between the density change and the Cauchy stress, because the average stress and the average density change are zero after the applied loads are totally removed. The level 3 condition is stringent and usually does not apply for the material whose elastic volume change is not linearly proportional to the stress.

Therefore in this paper we focus on how to satisfy level 2 condition, and it is what we mean by the term accurate volume conservation condition unless explicitly stated otherwise. It is also suggested using $\Delta\rho / \rho_{(0)}$ to measure the volume change due to the plastic deformation after the loads are removed, where $\rho_{(0)}$ is the density of the initial stress free state, $\Delta\rho$ is the difference between the density of state when the loads are totally removed and $\rho_{(0)}$, as schematically shown in Fig. 1. In this paper, we only concern volume conservation when the loads are removed (or "volume conservation" hereafter for simplicity). For the state under loading as denoted by the blue square in Fig. 1, the volume conservation is checked only on its corresponding unloading states denoted by the red circle.

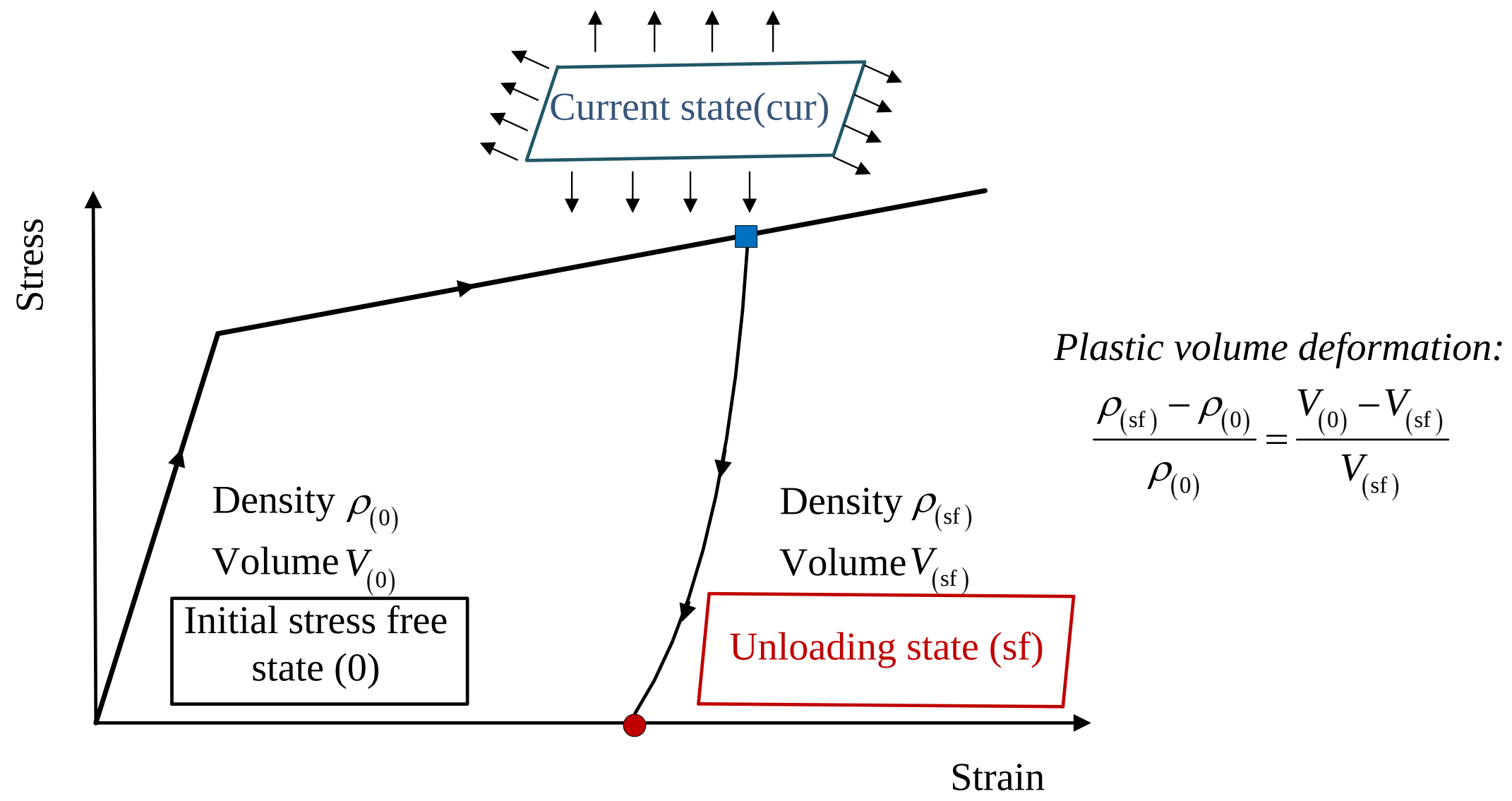

Fig. 1. Illustration of the accurate volume conservation condition and the definition of plastic volume deformation. The state whose plastic volume deformation is to be defined is denoted by a blue square, and its corresponding unloading stress free state is denoted by a red circle. The accurate volume conservation condition is $\rho_{(sf)} = \rho_{(0)}$.

## 3. Evaluation of classical theories on the accurate volume conservation condition

In the following, we will demonstrate or prove that many existing widely used elastoplastic constitutive theories and commercial software do not satisfy the proposed accurate volume conservation condition in the finite deformation regime. The tradition way in most theories to realize volume conservation is setting the plastic strain rate to be a deviatoric tensor. Considering that the plastic strain rate has many different definitions for finite deformation, obviously not all theories can realize the volume conservation, as discussed in more details in Section 3.1.

### 3.1. Various plastic strain rates defined by different theories

In establishing an elastoplastic theory at finite deformation, one usually has to make the following choices at least:

(1) The manner of strain rate decomposition. Among the various candidates, three ways suggested by three classical theories are considered in this paper, namely the Rice–Hill theory (Rice, 1971, 1975; Hill and Rice, 1972, 1973; Hill 1978), the Simo–Ortiz theory (Green and Naghdi, 1965, 1971; Simo and Ortiz, 1985) and the Moran–Ortiz–Shih theory (Moran et al., 1990). The plastic strain rate (or increment) of the first two theories are illustrated by the red segments in Fig.2.

(2) The reference configuration that is used to define the strain and the stress. Three configurations are usually used, which are the initial stress free configuration, the current configuration and the unloading stress free configuration, as shown in Fig 1.

(3) The stress/strain measure (See Appendix A for reference).

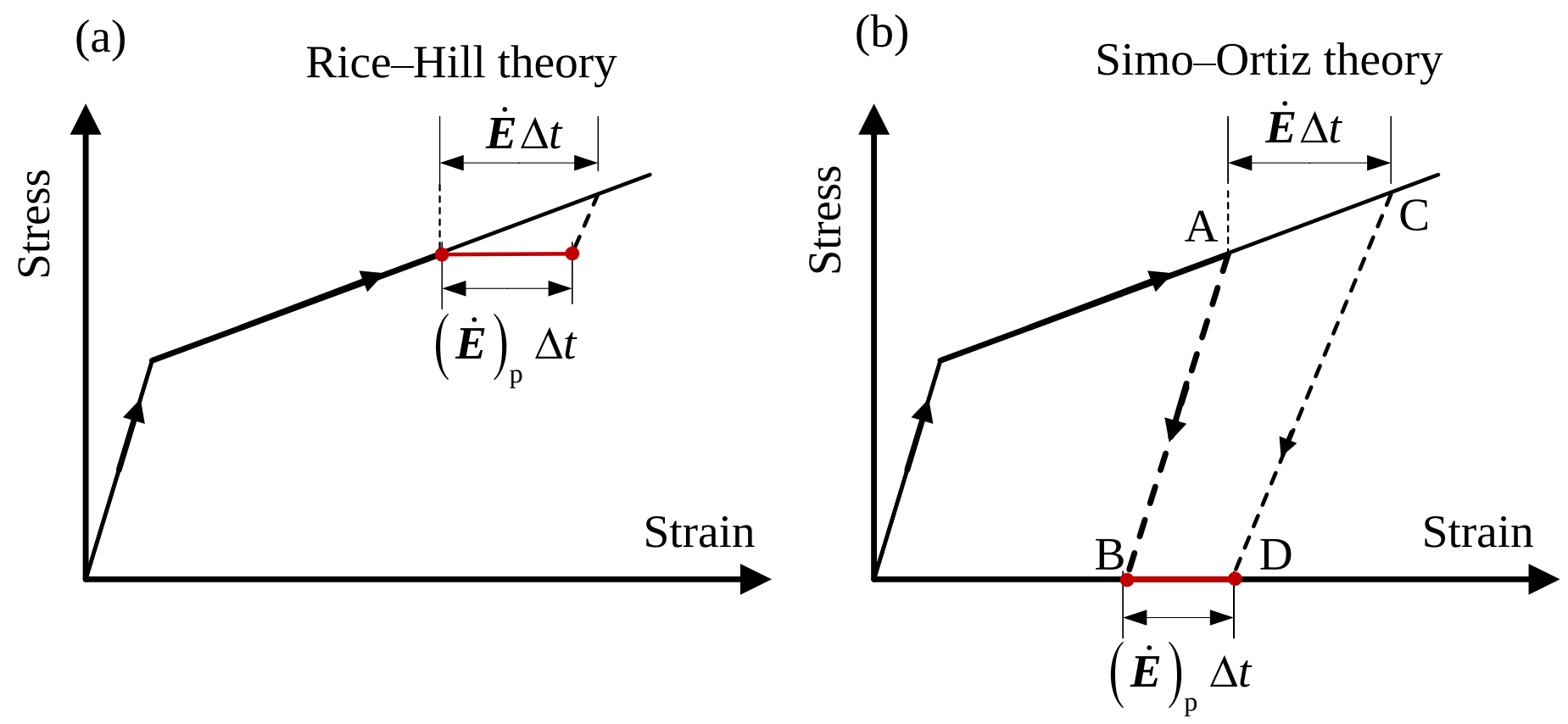

Fig. 2. The definitions of the plastic strain rate suggested by (a) the Rice–Hill theory and (b) the Simo–Ortiz theory.

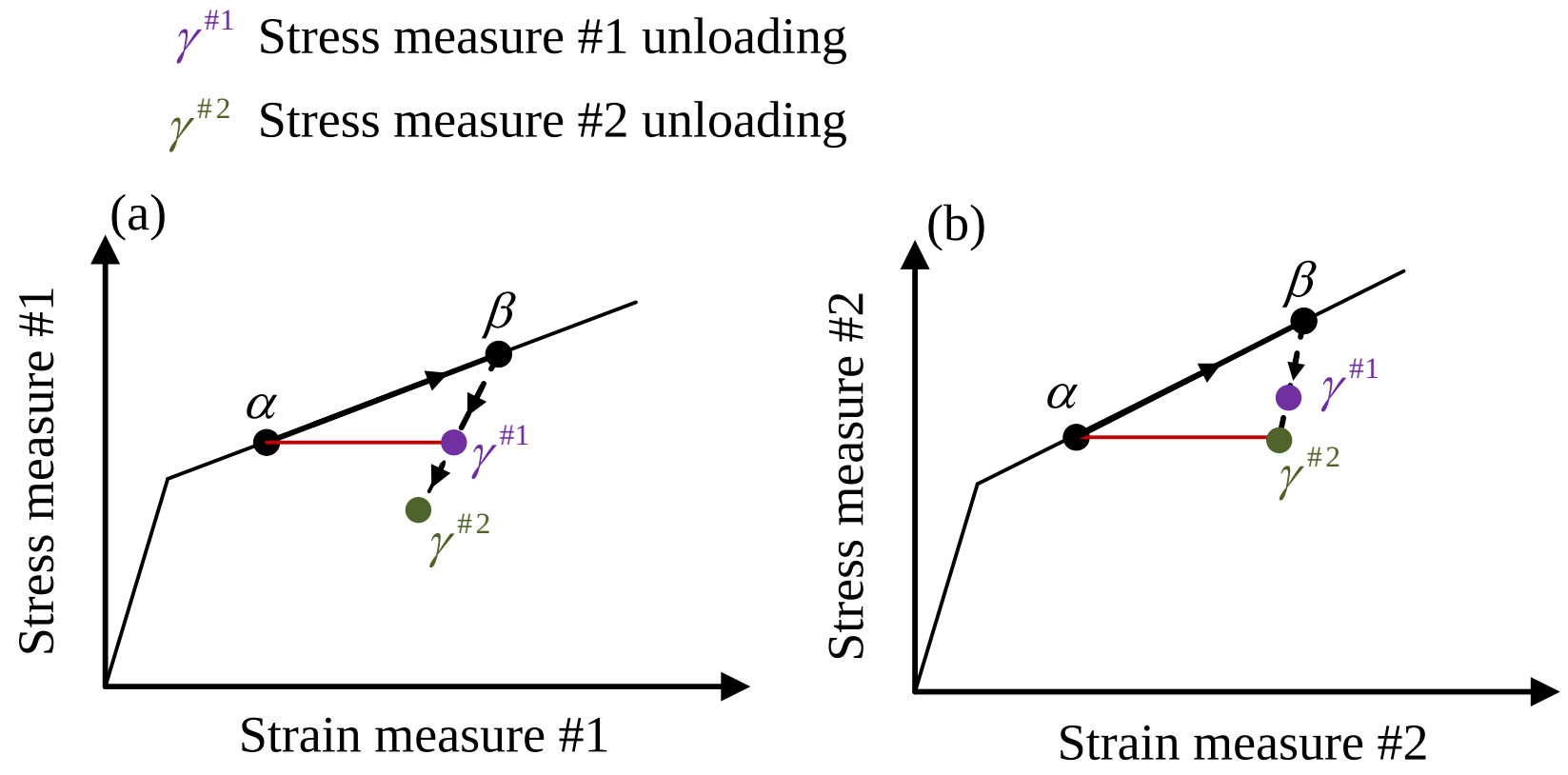

Fig. 3. An illustration that the strain rate decomposition suggested by the Rice–Hill theory is dependent on the stress/strain measure.

The above three choices can lead to many plastic strain rates. To illustrate how the arbitrary choice of the stress measure can affect the strain rate decomposition, the Rice–Hill theory with different stress unloading manners is discussed briefly here. Figure 3 schematically shows two stress–strain curves with two different stress/strain measures. Point $\alpha$ in Fig. 3(a) and Fig. 3(b) represent the same configuration, named as configuration $\alpha$, and the corresponding stresses are denoted by $\boldsymbol{\sigma}_{(\alpha)}^{\#1}$ and $\boldsymbol{\sigma}_{(\alpha)}^{\#2}$ for two stress measures respectively. After incremental deformation, the configuration becomes configuration $\beta$, denoted by Point $\beta$. In the strain rate decomposition suggested by the Rice–Hill theory, defining the plastic strain increment from configuration $\alpha$ to configuration $\beta$ requires the third configuration $\gamma$, which is achieved by unloading the stress to the same value of configuration $\alpha$. The plastic part of the deformation increment can then be defined by the difference of strain between configuration $\alpha$ and configuration $\gamma$. However, with different stress measures, e.g. stress measure #1 and #2, unloading the stress to the same value of configuration $\alpha$ will lead to different configuration $\gamma$, such as $\gamma^{\#1}$ and $\gamma^{\#2}$ in Fig. 3, and different plastic deformation increments (or plastic strain increments). It should be pointed out that the relative difference between these different plastic deformation increments does not vanish when the strain increment becomes infinitesimal, as demonstrated in the following example.

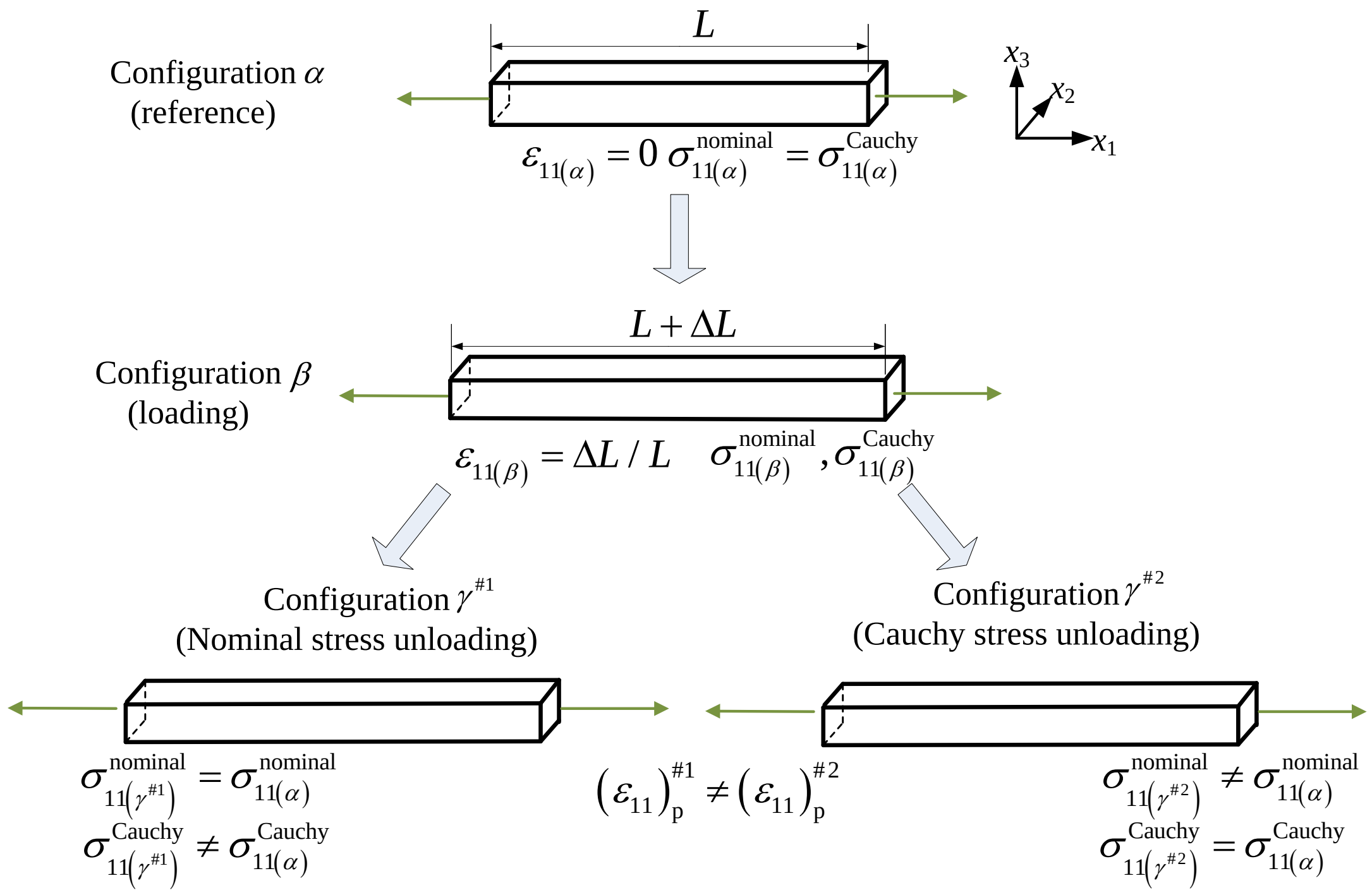

Fig. 4. A uniaxial tension example to illustrate that the strain decomposition suggested by the Rice–Hill theory depends on the stress/strain measure.

As shown in Fig. 4, a bar under uniaxial tension is investigated. Stress measure #1 and #2 are chosen to be the nominal stress and the Cauchy stress, respectively. For convenience we chose configuration $\alpha$ as the reference configuration so that the nominal stress and Cauchy stress at this moment are the same, i.e. $\sigma_{11(\alpha)}^{\text{nominal}} = \sigma_{11(\alpha)}^{\text{Cauchy}}$. Next we stretch the bar to configuration $\beta$, changing its length from $L$ to $L + \Delta L$, and the strain is

$$\varepsilon_{11(\beta)} = \frac{\Delta L}{L} \tag{1}$$

The stress increment can be determined from elastoplastic constitutive relation. Suppose the nominal stress increment during loading process is related to strain increment by

$$\Delta\sigma_{11}^{\text{nominal}} = C_{\text{ep}}\varepsilon_{11} \tag{2}$$

and during unloading process

$$\Delta\sigma_{11}^{\text{nominal}} = C_{\text{e}}\varepsilon_{11} \tag{3}$$

where $C_{\text{ep}}$ and $C_{\text{e}}$ are the tangential loading and unloading stiffness, respectively. So the nominal stress in configuration $\beta$ is

$$\sigma_{11(\beta)}^{\text{nominal}} = \sigma_{11(\alpha)}^{\text{nominal}} + C_{\text{ep}}\varepsilon_{11(\beta)} \tag{4}$$

and the Cauchy stress (or true stress) in configuration $\beta$ is determined as

$$\sigma_{11(\beta)}^{\text{Cauchy}} = \frac{A_{(\alpha)}}{A_{(\beta)}}\sigma_{11(\beta)}^{\text{nominal}} = \sigma_{11(\beta)}^{\text{nominal}}\left(1 + \varepsilon_{11(\beta)}\right) \tag{5}$$

where $A_{(\cdot)}$ indicates the area of the cross section in a certain configuration and volume conservation of deformation is assumed in obtained Eq. (5). If we unload the nominal stress to its previous value, namely $\sigma_{11(\gamma^{\#1})}^{\text{nominal}} = \sigma_{11(\alpha)}^{\text{nominal}}$, we obtain the nominal stress unloading configuration $\gamma^{\#1}$ and the plastic strain

$$\left(\varepsilon_{11}\right)_{\mathrm{p}}^{\#1}=\varepsilon_{11\left(\gamma^{\#1}\right)}=\varepsilon_{11(\beta)}-\frac{\sigma_{11(\beta)}^{\text{nominal}}-\sigma_{11(\alpha)}^{\text{nominal}}}{C_{\mathrm{e}}}=\frac{C_{\mathrm{e}}-C_{\mathrm{ep}}}{C_{\mathrm{e}}}\varepsilon_{11(\beta)} \tag{6}$$

and the Cauchy stress in configuration $\gamma^{\#1}$ is

$$\sigma_{11\left(\gamma^{\#1}\right)}^{\text{Cauchy}}=\sigma_{11\left(\gamma^{\#1}\right)}^{\text{nominal}}\left[1+\left(\varepsilon_{11}\right)_{\mathrm{p}}^{\#1}\right]=\sigma_{11(\alpha)}^{\text{nominal}}\left(1+\frac{C_{\mathrm{e}}-C_{ep}}{C_{\mathrm{e}}}\varepsilon_{11(\beta)}\right)>\sigma_{11(\alpha)}^{\text{Cauchy}}=\sigma_{11(\alpha)}^{\text{nominal}} \tag{7}$$

It is obvious that the Cauchy stress does not come back to its original value of configuration $\alpha$ when the nominal stress does. On the other hand, if we unload the Cauchy stress, namely $\sigma_{11\left(\gamma^{\#2}\right)}^{\text{Cauchy}}=\sigma_{11(\alpha)}^{\text{Cauchy}}=\sigma_{11(\alpha)}^{\text{nominal}}$, we obtain the Cauchy stress unloading configuration $\gamma^{\#2}$ and

$$\sigma_{11\left(\gamma^{\#2}\right)}^{\text{nominal}}=\frac{1}{1+\left(\varepsilon_{11}\right)_{\mathrm{p}}^{\#2}}\sigma_{11\left(\gamma^{\#2}\right)}^{\text{Cauchy}}=\frac{1}{1+\left(\varepsilon_{11}\right)_{\mathrm{p}}^{\#2}}\sigma_{11(\alpha)}^{\text{nominal}}<\sigma_{11(\alpha)}^{\text{nominal}}=\sigma_{11(\alpha)}^{\text{Cauchy}} \tag{8}$$

Therefore from

$$\sigma_{11\left(\gamma^{\#2}\right)}^{\text{nominal}}+C_{\mathrm{e}}\left[\varepsilon_{11(\beta)}-\varepsilon_{11\left(\gamma^{\#2}\right)}\right]=\sigma_{11(\beta)}^{\text{nominal}}=\sigma_{11(\alpha)}^{\text{nominal}}+C_{\mathrm{ep}}\varepsilon_{11(\beta)} \tag{9}$$

and Eq. (6), we obtain

$$\left(\varepsilon_{11}\right)_{\mathrm{p}}^{\#2}=\varepsilon_{11\left(\gamma^{\#2}\right)}=\frac{C_{\mathrm{e}}-C_{\mathrm{ep}}}{C_{\mathrm{e}}+\sigma_{11(\alpha)}^{\text{nominal}}}\varepsilon_{11(\beta)}=\frac{C_{\mathrm{e}}}{\left[C_{\mathrm{e}}+\sigma_{11(\alpha)}^{\text{nominal}}\right]}\left(\varepsilon_{11}\right)_{\mathrm{p}}^{\#1} \tag{10}$$

It is noted from Eq. (10) that $\left(\varepsilon_{11}\right)_{\mathrm{p}}^{\#1}$ and $\left(\varepsilon_{11}\right)_{\mathrm{p}}^{\#2}$ are unequal even though $\varepsilon_{11(\beta)}\to 0$. Taking $\sigma_{11(\alpha)}^{\text{nominal}}=\sigma_{11(\alpha)}^{\text{Cauchy}}=0.1C_{\mathrm{e}}$ and $C_{\mathrm{ep}}=0.1C_{\mathrm{e}}$, numerical result shows that the relative error of these two plastic strain increments (or rates) could be about 10%.

In summary, because there are so many different definitions of the plastic strain rate, setting all these different plastic strain rates to be deviatoric tensors cannot always realize volume conservation. A rigorous theoretical analysis is presented in the next subsection.

### 3.2. Theoretical evaluation on the volume conservation

In Section 2, the accurate condition of volume conservation is expressed by $\Delta\rho/\rho_{(0)}=0$. Based on this condition we evaluate some classical elastoplastic theories, as presented below. In most elastoplastic theories, it is the first and crucial step to define the plastic strain rate. According to whether the theories adopt the unloading stress free configuration or not in defining the plastic strain rate, we classify them into two categories in our discussion. Theories that adopt the unloading stress free configuration usually use the multiplicative decomposition and define both the plastic strain and the plastic strain rate. These theories have a relatively greater chance to realize volume conservation, as they always have the information of the unloading stress free configuration where the accurate condition of volume conservation is checked. But the use of the unloading stress free configuration is also their shortcoming. Other theories, such as the Rice–Hill theory (Rice, 1971, 1975; Hill and Rice, 1972, 1973; Hill 1978), do not need the unloading stress free configuration and define the plastic strain rate directly. These theories can hardly realize volume conservation by the tradition way of setting the plastic strain rate to be deviatoric. Among the large number of elastoplastic theories in literature, we choose the Simo–Ortiz theory and the Moran–Ortiz–Shih theory to represent the theories that adopt the unloading stress free configuration, and the Rice–Hill theory to represent those that do not adopt the unloading stress free configuration. With different configurations chosen as the reference configuration, the means to define the plastic strain rate adopted in these three theories are also used in many other theories, so we evaluate their volume conservations theoretically in Section 3.2 and numerically in Section 3.3. Furthermore the evaluation of a list of other theories is presented in Appendix B.

### 3.2.1. Theories that adopt the unloading stress free configuration

We first discuss the case when the unloading stress free configuration can be used. We denote the deformation gradient of the loading process (from the initial stress free configuration to the current configuration) and the unloading process (from the current configuration to the unloading stress free configuration) by $\boldsymbol{F}$ and $\left(\boldsymbol{F}^{\mathrm{e}}\right)^{-1}$ respectively. So the plastic deformation gradient from the initial stress free configuration to the unloading stress free configuration is

$$\boldsymbol{F}^{\mathrm{p}} = \left(\boldsymbol{F}^{\mathrm{e}}\right)^{-1} \cdot \boldsymbol{F} \tag{11}$$

where the dot represents the dot product of two tensors, namely $\boldsymbol{A} \cdot \boldsymbol{B} = A_{ik} B_{kj}$ .

The accurate condition of volume conservation can be written as

$$\frac{\rho_{(0)}}{\rho_{(\mathrm{sf})}} = \frac{V_{(\mathrm{sf})}}{V_{(0)}} = J^{\mathrm{p}} = \det(\boldsymbol{F}^{\mathrm{p}}) = 1 \tag{12}$$

where subscript (sf) and (0) indicate that the quantities are in the unloading stress free configuration and the initial stress free configuration, respectively; $\rho_{(\mathrm{sf})} = \rho_{(0)} + \Delta\rho$ is the density of material in the unloading stress free configuration; $V_{(\mathrm{sf})}$ and $V_{(0)}$ are the volumes of a material element; $J^{\mathrm{p}}$ is the volume ratio of the unloading stress free configuration. Taking derivatives on Eq. (12), we have

$$\dot{J}^{\mathrm{p}} = \frac{\partial \det(\boldsymbol{F}^{\mathrm{p}})}{\partial \boldsymbol{F}^{\mathrm{p}}} : \dot{\boldsymbol{F}}^{\mathrm{p}} = J^{\mathrm{p}} (\boldsymbol{F}^{\mathrm{p}})^{-\mathrm{T}} : \dot{\boldsymbol{F}}^{\mathrm{p}} = J^{\mathrm{p}} \mathrm{tr}\left( (\boldsymbol{F}^{\mathrm{p}})^{-1} \cdot \dot{\boldsymbol{F}}^{\mathrm{p}} \right) = J^{\mathrm{p}} \mathrm{tr}\left( \dot{\boldsymbol{F}}^{\mathrm{p}} \cdot (\boldsymbol{F}^{\mathrm{p}})^{-1} \right) = 0 \tag{13}$$

where the double dot represents the scalar product, namely $\boldsymbol{A} : \boldsymbol{B} = A_{ij} B_{ij}$ , $\det(\boldsymbol{F}^{\mathrm{p}})$ is the third invariant of $\boldsymbol{F}^{\mathrm{p}}$ and $\mathrm{tr}\left(\boldsymbol{A}\right) = A_{11} + A_{22} + A_{33}$ is the first invariant and called the trace of a second order tensor $\boldsymbol{A}$ . The accurate volume conservation condition is then reduced to

$$\mathrm{tr}\left( \dot{\boldsymbol{F}}^{\mathrm{p}} \cdot (\boldsymbol{F}^{\mathrm{p}})^{-1} \right) = 0 \tag{14}$$

Equation (14) is referred to in this paper as the unloading stress free form of volume conservation condition, since it needs the information of the unloading stress free configuration.

As mentioned in Section 3.1, there are at least three different choices on the reference configuration and three manners of strain rate decomposition, giving rise to at least nine different plastic strain rates. Six of them, namely those that use the Simo–Ortiz or the Moran–Ortiz–Shih strain rate decompositions, depend on the unloading stress free configuration. The existing elastoplastic theories usually set one of these plastic strain rates to be a deviatoric tensor, which will be checked if Eq. (14) is rigorously satisfied.

The plastic strain rates of the Simo–Ortiz strain rate decomposition are

$$\begin{cases} \left(\dot{\boldsymbol{E}}^{\mathrm{SO}}\right)_{\mathrm{p}} = \dfrac{1}{2}\left( \dot{\boldsymbol{F}}^{\mathrm{pT}} \cdot \boldsymbol{F}^{\mathrm{p}} + \boldsymbol{F}^{\mathrm{pT}} \cdot \dot{\boldsymbol{F}}^{\mathrm{p}} \right) \\ \left(\dot{\bar{\boldsymbol{E}}}^{\mathrm{SO}}\right)_{\mathrm{p}} = (\boldsymbol{F}^{\mathrm{p}})^{-\mathrm{T}} \cdot \left(\dot{\boldsymbol{E}}^{\mathrm{SO}}\right)_{\mathrm{p}} \cdot (\boldsymbol{F}^{\mathrm{p}})^{-1} = \dfrac{1}{2}\left[ (\boldsymbol{F}^{\mathrm{p}})^{-\mathrm{T}} \cdot \dot{\boldsymbol{F}}^{\mathrm{pT}} + \dot{\boldsymbol{F}}^{\mathrm{p}} \cdot (\boldsymbol{F}^{\mathrm{p}})^{-1} \right] \\ \left(\boldsymbol{d}^{\mathrm{SO}}\right)_{\mathrm{p}} = \boldsymbol{F}^{-\mathrm{T}} \cdot \left(\dot{\boldsymbol{E}}^{\mathrm{SO}}\right)_{\mathrm{p}} \cdot \boldsymbol{F}^{-1} = \dfrac{1}{2}\left[ (\boldsymbol{F}^{\mathrm{e}})^{-\mathrm{T}} \cdot (\boldsymbol{F}^{\mathrm{p}})^{-\mathrm{T}} \cdot \dot{\boldsymbol{F}}^{\mathrm{pT}} \cdot (\boldsymbol{F}^{\mathrm{e}})^{-1} + (\boldsymbol{F}^{\mathrm{e}})^{-\mathrm{T}} \cdot \dot{\boldsymbol{F}}^{\mathrm{p}} \cdot (\boldsymbol{F}^{\mathrm{p}})^{-1} \cdot (\boldsymbol{F}^{\mathrm{e}})^{-1} \right] \end{cases}$$

(15a-c)

where $\dot{\boldsymbol{E}}$ , $\dot{\bar{\boldsymbol{E}}}$ and $\boldsymbol{d}$ denote the strain rate when taking the initial stress free configuration, the unloading stress free configuration and the current configuration as the reference configuration, respectively. Here the strain measure is chosen to be the Green strain $\boldsymbol{E}$ . A superscript is added adjacent to the symbol to distinguish the manner of the strain rate decomposition, with SO standing for Simo–Ortiz, RH standing for Rice–Hill and MOS standing for Moran–Ortiz–Shih. The subscript "p" outside the bracket indicates the term is the plastic part, while subscript "e" indicates that the term is the elastic part. A combination of the choice on the manner of strain rate decomposition and the reference configuration gives rise to different elastoplastic theories. In abbreviation, the theories are named as SO-ini, SO-cur and SO-sf for the Simo–Ortiz strain rate decomposition with the initial stress free, current and the unloading stress free configuration as the reference configuration, respectively. SO can be replaced by either MOS or RH to denote theories using the other two strain rate decompositions.

The traces of the strain rates in Eq.(15) are

$$\begin{cases} \mathrm{tr}\left(\left(\dot{\boldsymbol{E}}^{\mathrm{SO}}\right)_{\mathrm{p}}\right)=\mathrm{tr}\left(\boldsymbol{F}^{\mathrm{pT}}\cdot\dot{\boldsymbol{F}}^{\mathrm{p}}\cdot(\boldsymbol{F}^{\mathrm{p}})^{-1}\cdot\boldsymbol{F}^{\mathrm{p}}\right) \\ \mathrm{tr}\left(\left(\dot{\bar{\boldsymbol{E}}}^{\mathrm{SO}}\right)_{\mathrm{p}}\right)=\mathrm{tr}\left(\dot{\boldsymbol{F}}^{\mathrm{p}}\cdot(\boldsymbol{F}^{\mathrm{p}})^{-1}\right) \\ \mathrm{tr}\left(\left(\boldsymbol{d}^{\mathrm{SO}}\right)_{\mathrm{p}}\right)=\mathrm{tr}\left(\left(\boldsymbol{F}^{\mathrm{e}}\right)^{-\mathrm{T}}\cdot\dot{\boldsymbol{F}}^{\mathrm{p}}\cdot\left(\boldsymbol{F}^{\mathrm{p}}\right)^{-1}\cdot\left(\boldsymbol{F}^{\mathrm{e}}\right)^{-1}\right) \end{cases} \tag{16a-c}$$

For the Moran–Ortiz–Shih theory, we have

$$\begin{cases} \left(\dot{\boldsymbol{E}}^{\mathrm{MOS}}\right)_{\mathrm{p}}=\frac{1}{2}\left(\boldsymbol{F}^{\mathrm{pT}}\cdot\boldsymbol{F}^{\mathrm{eT}}\cdot\boldsymbol{F}^{\mathrm{e}}\cdot\dot{\boldsymbol{F}}^{\mathrm{p}}+\dot{\boldsymbol{F}}^{\mathrm{pT}}\cdot\boldsymbol{F}^{\mathrm{eT}}\cdot\boldsymbol{F}^{\mathrm{e}}\cdot\boldsymbol{F}^{\mathrm{p}}\right) \\ \left(\dot{\bar{\boldsymbol{E}}}^{\mathrm{MOS}}\right)_{\mathrm{p}}=(\boldsymbol{F}^{\mathrm{p}})^{-\mathrm{T}}\cdot\left(\dot{\boldsymbol{E}}^{\mathrm{MOS}}\right)_{\mathrm{p}}\cdot(\boldsymbol{F}^{\mathrm{p}})^{-1}=\frac{1}{2}\left[\boldsymbol{F}^{\mathrm{eT}}\cdot\boldsymbol{F}^{\mathrm{e}}\cdot\dot{\boldsymbol{F}}^{\mathrm{p}}\cdot(\boldsymbol{F}^{\mathrm{p}})^{-1}+(\boldsymbol{F}^{\mathrm{p}})^{-\mathrm{T}}\cdot\dot{\boldsymbol{F}}^{\mathrm{pT}}\cdot\boldsymbol{F}^{\mathrm{eT}}\cdot\boldsymbol{F}^{\mathrm{e}}\right] \\ \left(\boldsymbol{d}^{\mathrm{MOS}}\right)_{\mathrm{p}}=\boldsymbol{F}^{-\mathrm{T}}\cdot\left(\dot{\boldsymbol{E}}^{\mathrm{MOS}}\right)_{\mathrm{p}}\cdot\boldsymbol{F}^{-1}=\frac{1}{2}\left[\boldsymbol{F}^{\mathrm{e}}\cdot\dot{\boldsymbol{F}}^{\mathrm{p}}\cdot(\boldsymbol{F}^{\mathrm{p}})^{-1}\cdot(\boldsymbol{F}^{\mathrm{e}})^{-1}+(\boldsymbol{F}^{\mathrm{e}})^{-\mathrm{T}}\cdot(\boldsymbol{F}^{\mathrm{p}})^{-\mathrm{T}}\cdot\dot{\boldsymbol{F}}^{\mathrm{pT}}\cdot\boldsymbol{F}^{\mathrm{eT}}\right] \end{cases} \tag{17a-c}$$

$$\begin{cases} \mathrm{tr}\left(\left(\dot{\boldsymbol{E}}^{\mathrm{MOS}}\right)_{\mathrm{p}}\right)=\mathrm{tr}\left((\boldsymbol{F}^{\mathrm{p}})^{\mathrm{T}}\cdot\boldsymbol{F}^{\mathrm{eT}}\cdot\boldsymbol{F}^{\mathrm{e}}\cdot\dot{\boldsymbol{F}}^{\mathrm{p}}\cdot\left(\boldsymbol{F}^{\mathrm{p}}\right)^{-1}\cdot\boldsymbol{F}^{\mathrm{p}}\right) \\ \mathrm{tr}\left(\left(\dot{\bar{\boldsymbol{E}}}^{\mathrm{MOS}}\right)_{\mathrm{p}}\right)=\mathrm{tr}\left(\boldsymbol{F}^{\mathrm{e}}\cdot\dot{\boldsymbol{F}}^{\mathrm{p}}\cdot\left(\boldsymbol{F}^{\mathrm{p}}\right)^{-1}\cdot\boldsymbol{F}^{\mathrm{eT}}\right) \\ \mathrm{tr}\left(\left(\boldsymbol{d}^{\mathrm{MOS}}\right)_{\mathrm{p}}\right)=\mathrm{tr}\left(\dot{\boldsymbol{F}}^{\mathrm{p}}\cdot\left(\boldsymbol{F}^{\mathrm{p}}\right)^{-1}\right) \end{cases} \tag{18a-c}$$

The traces of the plastic strain rates derived above are also summarized in Table 1. It can be concluded by observing Eq. (16) and Eq. (18) that only $\mathrm{tr}\left(\left(\dot{\bar{\boldsymbol{E}}}^{\mathrm{SO}}\right)_{\mathrm{p}}\right)=0$ and $\mathrm{tr}\left(\left(\boldsymbol{d}^{\mathrm{MOS}}\right)_{\mathrm{p}}\right)=0$ satisfy the accurate volume conservation condition Eq. (14), namely only the SO-sf theory (Simo–Ortiz strain rate decomposition with the unloading stress free configuration as the reference configuration) and the MOS-cur theory (Moran–Ortiz–Shin strain rate decomposition with the current configuration as the reference configuration) are capable of rigorously realizing volume conservation. Therefore even the theories that adopt the unloading stress free configuration should be very careful to choose the proper plastic strain tensor to satisfy the accurate condition of volume conservation.

Table 1. Volume conservation evaluation of the elastoplastic theories using different strain decompositions suggested by the Rice–Hill, Simo–Ortiz and Moran–Ortiz–Shih theory with three different choices of the reference configurations. The symbol √ indicates that the volume conservation condition is satisfied by setting this term to be a deviatoric tensor, and × indicates the condition cannot be satisfied in this way.

| | Reference configuration | | |
|---|---|---|---|
| | Initial stress free configuration | Current configuration | Unloading stress free configuration |
| Rice–Hill | $\mathrm{tr}\left(\left(\dot{\boldsymbol{E}}^{\mathrm{RH}}\right)_{\mathrm{p}}\right)$ × | $\mathrm{tr}\left(\left(\boldsymbol{d}^{\mathrm{RH}}\right)_{\mathrm{p}}\right)$ × | $\mathrm{tr}\left(\left(\dot{\bar{\boldsymbol{E}}}^{\mathrm{RH}}\right)_{\mathrm{p}}\right)$ × |
| Simo–Ortiz | $\mathrm{tr}\left(\left(\dot{\boldsymbol{E}}^{\mathrm{SO}}\right)_{\mathrm{p}}\right)= \mathrm{tr}\left(\boldsymbol{F}^{\mathrm{pT}}\cdot\dot{\boldsymbol{F}}^{\mathrm{p}}\cdot(\boldsymbol{F}^{\mathrm{p}})^{-1}\cdot\boldsymbol{F}^{\mathrm{p}}\right)$ × | $\mathrm{tr}\left(\left(\boldsymbol{d}^{\mathrm{SO}}\right)_{\mathrm{p}}\right)= \mathrm{tr}\left(\left(\boldsymbol{F}^{\mathrm{e}}\right)^{-\mathrm{T}}\cdot\dot{\boldsymbol{F}}^{\mathrm{p}}\cdot\left(\boldsymbol{F}^{\mathrm{p}}\right)^{-1}\cdot\left(\boldsymbol{F}^{\mathrm{e}}\right)^{-1}\right)$ × | $\mathrm{tr}\left(\left(\dot{\bar{\boldsymbol{E}}}^{\mathrm{SO}}\right)_{\mathrm{p}}\right)= \mathrm{tr}\left(\dot{\boldsymbol{F}}^{\mathrm{p}}\cdot(\boldsymbol{F}^{\mathrm{p}})^{-1}\right)$ √ |
| Moran–Ortiz–Shih | $\mathrm{tr}\left(\left(\dot{\boldsymbol{E}}^{\mathrm{MOS}}\right)_{\mathrm{p}}\right)= \mathrm{tr}\left((\boldsymbol{F}^{\mathrm{p}})^{\mathrm{T}}\cdot\boldsymbol{F}^{\mathrm{eT}}\cdot\boldsymbol{F}^{\mathrm{e}}\cdot\dot{\boldsymbol{F}}^{\mathrm{p}}\cdot\left(\boldsymbol{F}^{\mathrm{p}}\right)^{-1}\cdot\boldsymbol{F}^{\mathrm{p}}\right)$ × | $\mathrm{tr}\left(\left(\boldsymbol{d}^{\mathrm{MOS}}\right)_{\mathrm{p}}\right)= \mathrm{tr}\left(\dot{\boldsymbol{F}}^{\mathrm{p}}\cdot\left(\boldsymbol{F}^{\mathrm{p}}\right)^{-1}\right)$ √ | $\mathrm{tr}\left(\left(\dot{\bar{\boldsymbol{E}}}^{\mathrm{MOS}}\right)_{\mathrm{p}}\right)= \mathrm{tr}\left(\boldsymbol{F}^{\mathrm{e}}\cdot\dot{\boldsymbol{F}}^{\mathrm{p}}\cdot\left(\boldsymbol{F}^{\mathrm{p}}\right)^{-1}\cdot\boldsymbol{F}^{\mathrm{eT}}\right)$ × |

### 3.2.2. Theories that do not adopt the unloading stress free configuration

For the theories that do not use the unloading stress free configuration such as the Rice–Hill theory, it is difficult to express the trace of the plastic strain rate by $\boldsymbol{F}^{\mathrm{p}}$, $\boldsymbol{F}^{\mathrm{e}}$, $\dot{\boldsymbol{F}}^{\mathrm{p}}$ and $\dot{\boldsymbol{F}}^{\mathrm{e}}$, but we can demonstrate that in general (see Appendix C)

$$\begin{cases} \mathrm{tr}\left(\left(\dot{\boldsymbol{E}}^{\mathrm{RH}}\right)_{\mathrm{p}}\right)\neq \mathrm{tr}\left(\left(\dot{\boldsymbol{E}}^{\mathrm{SO}}\right)_{\mathrm{p}}\right) \\ \mathrm{tr}\left(\left(\dot{\bar{\boldsymbol{E}}}^{\mathrm{RH}}\right)_{\mathrm{p}}\right)\neq \mathrm{tr}\left(\left(\dot{\bar{\boldsymbol{E}}}^{\mathrm{SO}}\right)_{\mathrm{p}}\right) \\ \mathrm{tr}\left(\left(\boldsymbol{d}^{\mathrm{RH}}\right)_{\mathrm{p}}\right)\neq \mathrm{tr}\left(\left(\boldsymbol{d}^{\mathrm{SO}}\right)_{\mathrm{p}}\right) \end{cases} \tag{19a-c}$$

Therefore the Rice–Hill theory usually cannot rigorously realize volume conservation by setting the plastic strain rate to be deviatoric no matter which configuration is chosen as the reference configuration.

## 3.3. Numerical evaluation on the volume conservation

A uniaxial loading example (Fig. 5(a)) is presented in this subsection to illustrate that when the accurate condition of volume conservation is not satisfied, significant errors will arise. The material is assumed to be isotropic linear strain hardening expressed by the relationship between the logarithmic strain and the Cauchy stress. The Young's modulus, Poisson's ratio and hardening coefficient of the material are denoted by $C_{\mathrm{e}}$, $\nu$ and $C_{\mathrm{ep}}$ respectively. Under uniaxial loading in the $x_1$ direction, the Cauchy stress $\sigma_{11}$ and the logarithmic strain $E_{11}^{\ln}$ are assumed to have the following relationship (Fig. 5(b))

$$\sigma_{11}=\begin{cases} C_{\mathrm{e}}E_{11}^{\ln} & 0\le E_{11}^{\ln}\le E_{\mathrm{cr}} \\ \sigma_{\mathrm{cr}}+C_{\mathrm{ep}}\left(E_{11}^{\ln}-E_{\mathrm{cr}}\right) & E_{11}^{\ln}>E_{\mathrm{cr}} \end{cases} \quad \text{for tension} \tag{20}$$

$$\sigma_{11}=\begin{cases} C_{\mathrm{e}}E_{11}^{\ln} & -E_{\mathrm{cr}}\le E_{11}^{\ln}<0 \\ -\sigma_{\mathrm{cr}}+C_{\mathrm{ep}}\left(E_{11}^{\ln}+E_{\mathrm{cr}}\right) & E_{11}^{\ln}<-E_{\mathrm{cr}} \end{cases} \quad \text{for compression} \tag{21}$$

where $\sigma_{\mathrm{cr}}$ is the initial yielding stress and $E_{\mathrm{cr}}=\sigma_{\mathrm{cr}}/C_{\mathrm{e}}$ is the initial yielding strain. Cartesian coordinate system is used here, and 1, 2, 3 stand for $x_1$, $x_2$, $x_3$, respectively. The elastic property of the material is assumed to be uncoupled with the plastic deformation, so under elastic uniaxial loading or unloading

$$\dot{E}_{11}^{\ln} = \frac{1}{C_{\mathrm{e}}}\dot{\sigma}_{11}, \dot{E}_{22}^{\ln} = \dot{E}_{33}^{\ln} = -\frac{\nu}{C_{\mathrm{e}}}\dot{\sigma}_{11} \tag{22}$$

It should be pointed out that the stress–strain relations Eq. (20) and Eq. (21) are expressed in terms of the Cauchy stress and the logarithmic strain, as required by the commercial software ABAQUS (version 6.14), since the plastic volume deformation predicted by ABAQUS is also evaluated in this paper. In the following numerical examples, the material parameters are set as $C_{\mathrm{ep}} = 0.1C_{\mathrm{e}}$, $\sigma_{\mathrm{cr}} = 0.001C_{\mathrm{e}}$ and $\nu = 0.3$

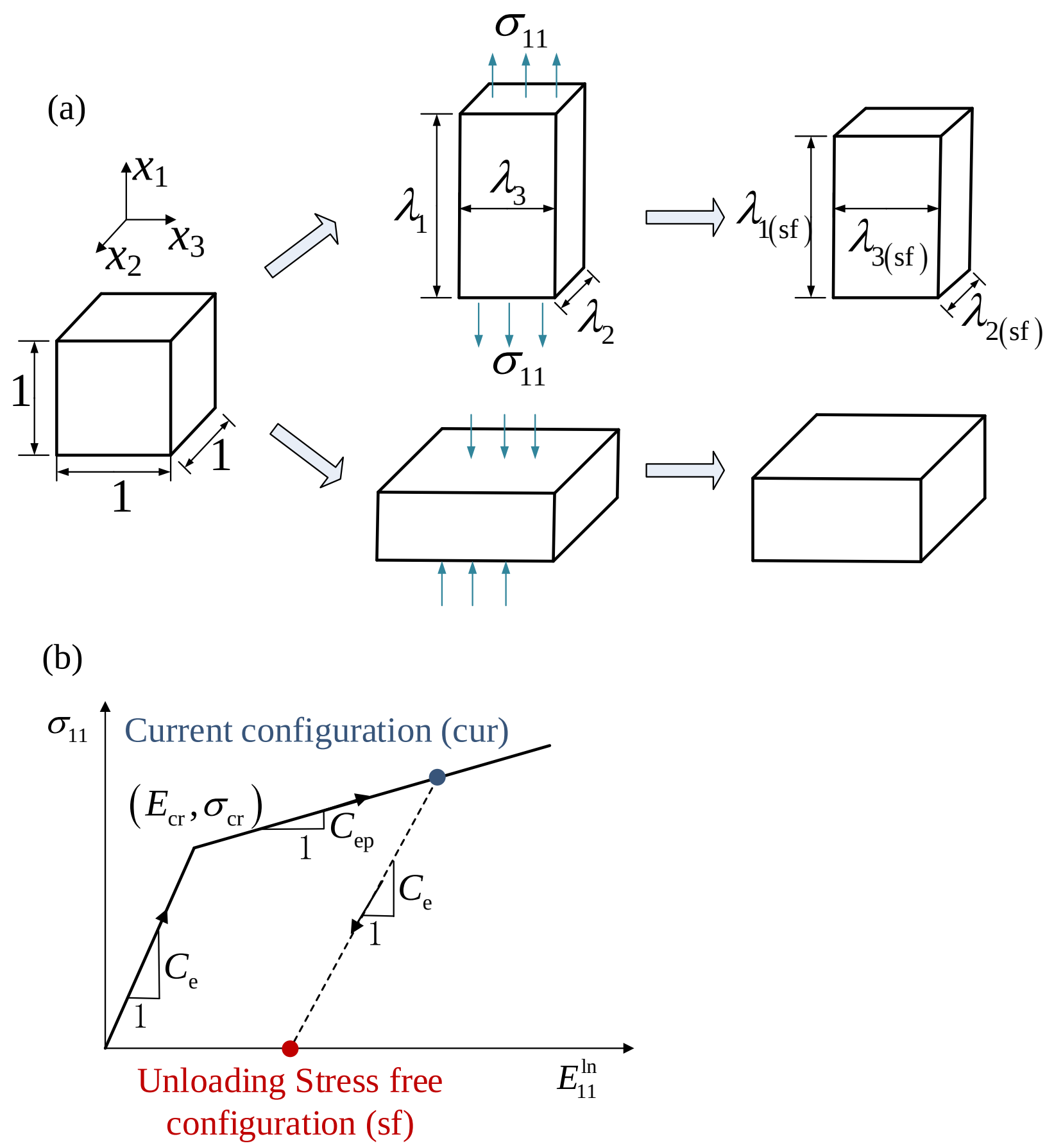

Fig. 5. (a) Illustration of the uniaxial loading example; (b) the stress–strain curve used in this example.

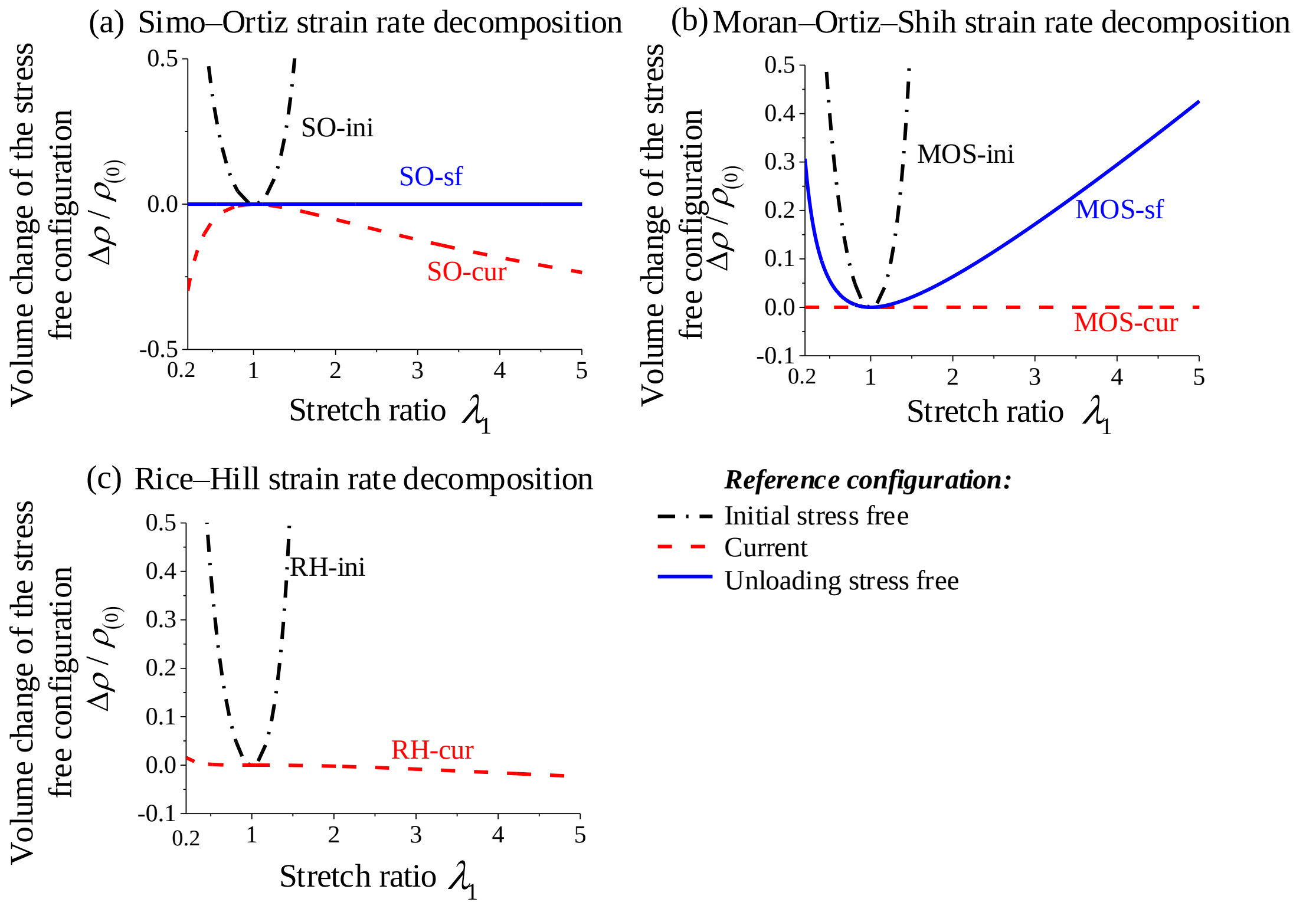

Fig. 6. The volume change of the unloading stress free configuration predicted by various elastoplastic theories.

#### 3.3.1. Plastic volume deformation

The plastic volume deformation is calculated as

$$\frac{\Delta\rho}{\rho_{(0)}}=\frac{\rho_{(\mathrm{sf})}-\rho_{(0)}}{\rho_{(0)}}=\frac{V_{(0)}-V_{(\mathrm{sf})}}{V_{(\mathrm{sf})}}=\left(\lambda_{1(\mathrm{sf})}\lambda_{2(\mathrm{sf})}\lambda_{3(\mathrm{sf})}\right)^{-1}-1 \tag{23}$$

where $\lambda_{i(\mathrm{sf})}$ is the stretch ratio in the unloading stress free configuration. The detailed formulae for calculating the plastic volume deformation in this uniaxial loading example are presented in Appendix D. The volume change of the unloading stress free configuration as a function of the stretch ratio of the loading configuration $\lambda_1$ is shown in Fig. 6 for different choices on the reference configuration and different manners of strain rate decompositions suggested by the Rice–Hill, Simo–Ortiz and Moran–Ortiz–Shih theories respectively. Only in two cases, namely the Simo–Ortiz strain rate decomposition with the unloading stress free configuration as the reference configuration (SO-sf) and the Moran–Ortiz–Shih strain decomposition with the current configuration as the reference configuration (MOS-cur), the volume conservation is realized.

The predicted plastic volume deformations for other theories presented in Fig. 6 (SO-ini, SO-cur, MOS-ini, MOS-sf, RH-ini and RH-cur) are not zero, and can be very large for the theories taking the initial configuration as the reference configuration (SO-ini, MOS-ini and RH-ini) in both compression and tension. The plastic volume deformations predicted by SO-ini, MOS-ini and RH-ini exceed 50% when the stretch ratio $\lambda_1=2$ (stretch to 2 times) or $\lambda_1 = 0.2$ (compress to 1/5). Among the evaluated theories that are not volume conserved, RH-cur has the best performance, but still predicts a plastic volume deformation of 3% when $\lambda_1 = 5$ or $\lambda_1 = 0.2$. The above results are obtained through numerical integral whose convergence is guaranteed by using a very small increment. Taking the SO-cur theory for example, the plastic volume deformation predicted by using different increments in the numerical integral is shown in Fig. 7. The calculation is convergent when the increment is reduced to $\lambda_{1t+\Delta t} / \lambda_{1t} = \mathrm{e}^{0.001}$, so we use increment smaller than this value in the numerical calculation.

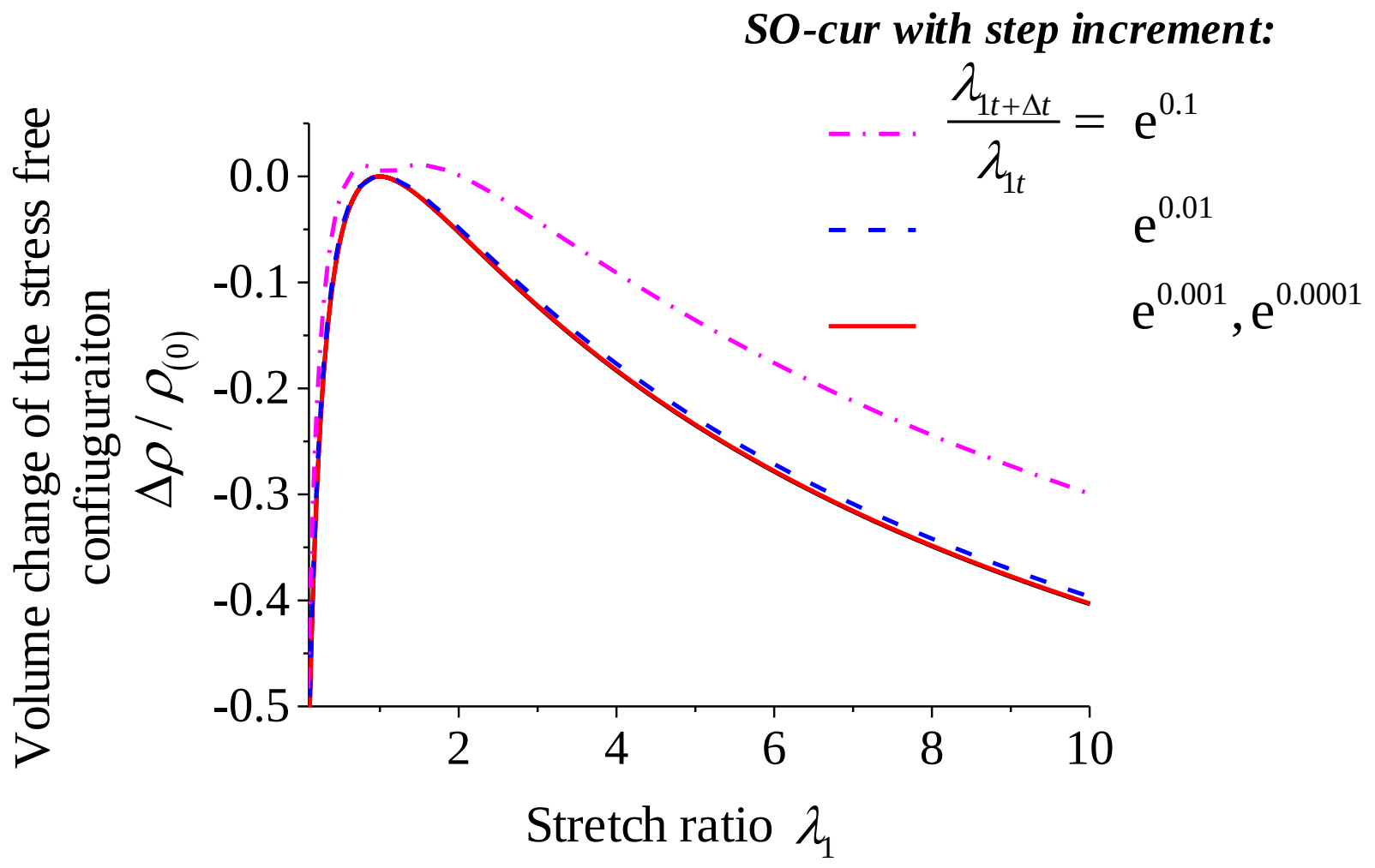

Fig. 7. An illustration of the numerical convergence in the uniaxial loading example

Finite element simulations are also carried out using commercial software ABAQUS (standard, version 6.14) , COMSOL (version 5.0) and ANSYS (version 16.0) respectively, with an 8-node brick element. Because the problem is nonlinear, the simulation is divided into many increments. The amount of stretch $\Delta\lambda_1$ applied in each increment affects the result of ABAQUS. Using a small $\Delta\lambda_1$, ABAQUS predicts almost zero plastic volume deformation but the simulation does not converge when the total applied stretch $\lambda_1$ is large (the black solid line in Fig. 8(a)). The convergence can be improved by increasing the increment $\Delta\lambda_1$, but the predicted plastic volume deformation also increases (the red dash-dotted line and the blue short dashed line in Fig. 8(a)). In brief the elastoplastic simulation of ABAQUS is not always volume conserved.

As shown by Fig. 8(b), COMSOL and ANSYS predict large plastic volume deformation even for quite small applied stretch ratio. Therefore they also do not predict volume conserved result for the elastoplastic problem of finite deformation.

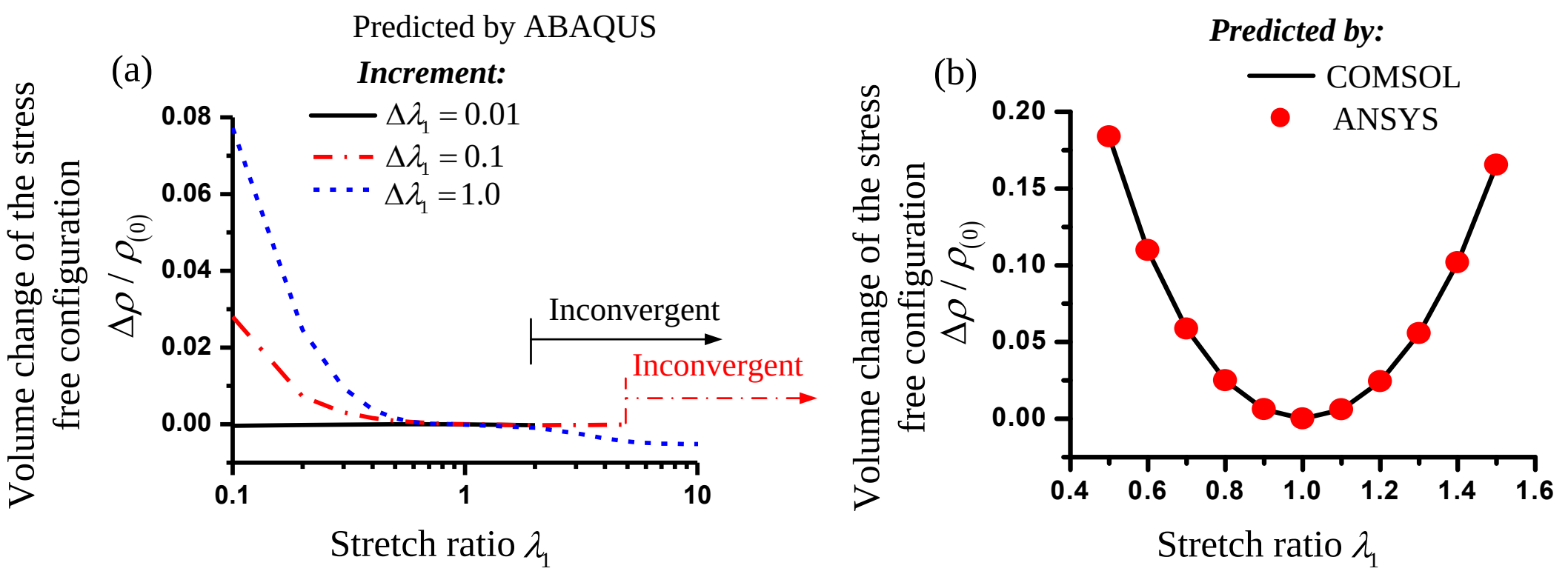

Fig. 8. The plastic volume deformation predicted by commercial software (a) ABAQUS; (b) COMSOL and ANSYS.

#### 3.3.2. Error on the loading stress

When an elastoplastic theory that cannot rigorously realize volume conservation is used, significant error could arise in the prediction of stress. Similar to the predictions on the plastic volume deformation, only SO-sf and MOS-cur theories predict the accurate stress. The theories using the initial configuration as the reference configuration give rise to a relative error larger than 50% when the required stretch ratio $\lambda_{2(\mathrm{sf})} > 1.3$ or $\lambda_{2(\mathrm{sf})} < 0.7$. The relative error of SO-cur and MOS-sf exceeds 20% when $\lambda_{2(\mathrm{sf})} > 3$ or $\lambda_{2(\mathrm{sf})} < 0.5$. SO-cur has the best performance, with an relative error about 1% for $0.5 < \lambda_{2(\mathrm{sf})} \le 3$.The above results are shown in Fig. E1 in Appendix E.

## 4. Strategies to realize volume conservation in elastoplastic constitutive theories.

### 4.1. Strategy 0: use theories such as SO-sf and MOS-cur when the unloading stress free configuration can be uniquely determined

If an elastoplastic theory satisfies the unloading stress free form of volume conservation condition Eq. (14), it is volume conserved. Among the theories evaluated in Section 3.2 and Section 3.3, by setting the plastic strain rate to be a deviatoric tensor only the SO-sf theory (Simo–Ortiz strain rate decomposition with the unloading stress free configuration as the reference configuration) and the MOS-cur theory (Moran–Ortiz–Shih strain rate decomposition with the current configuration as the reference configuration) realize volume conservation. There are also several theories in literature that realize the volume conservation via this strategy. But a major drawback for adopting this strategy is that the unloading stress free configuration cannot be avoided. For example in the SO-sf theory and the MOS-cur theory the plastic strain rates are defined through the plastic deformation gradient $\boldsymbol{F}^{\mathrm{p}}$ explicitly as illustrated by Eq. (15b) and Eq. (17c) , so they utilize the unloading stress free configurations and have many assumptions on the unloading process, such as

(1) No reverse plastic deformation occurs when the material is unloaded to the stress free configuration. This assumption is not always valid for materials showing kinetic hardening behavior.

(2) Usually the elastic moduli during unloading is assumed to be constant and the same as those during the initial loading process before any plastic deformation occurs, but this assumption is probably improper in large deformation because the elastic moduli might be intrinsically changed or extrinsically changed due to the measure dependence.

Above assumptions are sometimes too strong and cannot always be met, making the adoption of the unloading stress free configuration in elastoplastic theories a controversial issue. Therefore we should seek alternate strategies to establish elastoplastic theories which do not need to use the unloading stress free configuration so that most above assumptions can be discarded and the applicability is expanded. In such theories the unloading stress free form of the volume conservation condition Eq.(14) is difficult to satisfy directly because we have no information of the unloading stress free configuration. Therefore first of all, a condition equivalent to the accurate volume conservation condition should be proposed and expressed in the current configuration. Here a volume conservation condition is proposed based on only one assumption that the density of material is a function of the trace of the Cauchy stress, namely

$$\frac{\Delta\rho}{\rho_{(0)}} = \frac{1}{J} - 1 = g\left(\mathrm{tr}\left(\boldsymbol{\sigma}\right)\right) \tag{24}$$

where function $g(x)$ satisfies $g(0)=0$ and $g'(x)<0$. It is not difficult to determine $g(x)$ from experiment such as uniaxial tests. If Eq. (24) is satisfied, the volume conservation is realized because in the unloading stress free configuration $\boldsymbol{\sigma}=0$, giving rise to $\Delta\rho=0$. The reason to choose Cauchy stress in the above volume conservation condition is that Cauchy stress has a straightforward physical meaning without requiring a reference configuration and then has a unique position superior to other stress measures. The following discussion is mainly based on the Cauchy stress if no explicit indication is given. Hereafter Eq. (24) is referred to as the current deformation form of the volume conservation condition, and $g\left(\mathrm{tr}\left(\boldsymbol{\sigma}\right)\right)$ is called the volume constitutive function.

As mentioned previously that elastoplastic theories usually assume the trace of the plastic strain rate to be zero, and fail most times in ensuring the volume conservation. In the following, we will demonstrate that the volume conservation, however, corresponds to the zero trace of the plastic stress rate.

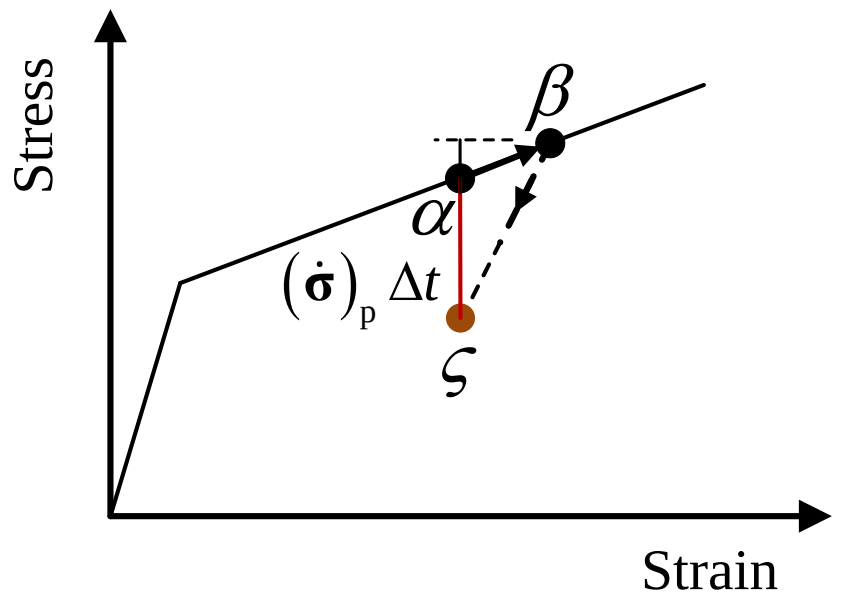

Fig. 9. Illustration of the stress rate decomposition and the plastic stress rate.

Figure 9 schematically shows the definition of the plastic stress rate $\left(\dot{\boldsymbol{\sigma}}\right)_{\mathrm{p}}$. When the material deforms from the current configuration $\alpha$ with stress $\boldsymbol{\sigma}_{(\alpha)}$ to a nearby configuration $\beta$ with stress $\boldsymbol{\sigma}_{(\beta)}$, and then is unloaded to configuration $\varsigma$ whose deformation gradient $\boldsymbol{F}$ is the same as that of configuration $\alpha$, the stress difference between configuration $\alpha$ and configuration $\varsigma$ is defined as the plastic stress increment $\left(\dot{\boldsymbol{\sigma}}\right)_{\mathrm{p}}\Delta t = \boldsymbol{\sigma}_{(\alpha)} - \boldsymbol{\sigma}_{(\varsigma)}$, and the plastic stress rate is defined as $\left(\dot{\boldsymbol{\sigma}}\right)_{\mathrm{p}} = \lim_{\Delta t\to 0}\left(\boldsymbol{\sigma}_{(\alpha)} - \boldsymbol{\sigma}_{(\varsigma)}\right)/\Delta t$. In general an objective rate should be used when taking the derivative of the Cauchy stress with respect to time, but the plastic stress rate defined above is already an objective tensor as configuration $\alpha$ and $\varsigma$ have the same deformation gradient so all the objective rates converge to the material rate.

It should be emphasized that the strain unloading configuration $\varsigma$ is unique, because the identical deformation gradient has unambiguous meaning, while the stress unloading configuration has many different choices corresponding to different stress measures as shown in Section 3.1 (Fig. 3 and Fig. 4). Therefore the plastic stress rate can be defined uniquely but the plastic strain rate cannot.

From Eq.(24), we obtain

$$\frac{\rho_{(\varsigma)}-\rho_{(\alpha)}}{\rho_{(0)}}=g\left(\mathrm{tr}\left(\boldsymbol{\sigma}_{(\alpha)}\right)-\mathrm{tr}\left(\left(\dot{\boldsymbol{\sigma}}\right)_{\mathrm{p}}\Delta t\right)\right)-g\left(\mathrm{tr}\left(\boldsymbol{\sigma}_{(\alpha)}\right)\right)=-g'\left(\mathrm{tr}\left(\boldsymbol{\sigma}_{(\alpha)}\right)\right)\mathrm{tr}\left(\left(\dot{\boldsymbol{\sigma}}\right)_{\mathrm{p}}\right)\Delta t \quad (25)$$

The deformations of configuration $\varsigma$ and $\alpha$ are the same, so $\rho_{(\varsigma)}=\rho_{(\alpha)}$ and a very concise volume conservation condition in rate form is

$$\mathrm{tr}\left(\dot{\boldsymbol{\sigma}}\right)_{\mathrm{p}}=0 \quad (26)$$

It is interesting to note that the zero trace of the plastic stress rate tensor, not the plastic strain rate tensor, is equivalent to the accurate condition of volume conservation.

In the following parts of this section, we propose two new strategies to realize volume conservation based on Eq. (24).

4.2. New strategy 1 directly and slightly revising the tangential stiffness tensor

Equation (24) provides a volume conservation condition in deformation form to constrain the elastoplastic theory, and we begin this subsection by deriving a rate form of it.

Taking derivatives on Eq. (24) yields

$$-\frac{\dot{J}}{J^2}=g'\left(\mathrm{tr}\left(\boldsymbol{\sigma}\right)\right)\mathrm{tr}\left(\dot{\boldsymbol{\sigma}}\right) \quad (27)$$

On the other hand, by noticing that $J=\det\left(\boldsymbol{F}\right)$, we can derive

$$\dot{J}=J\mathrm{tr}\left(\boldsymbol{d}\right) \quad (28)$$

Substituting Eq. (28) into Eq. (27) yields another volume conservation condition in rate form

$$\mathrm{tr}\left(\boldsymbol{d}\right)=-Jg'\left(\mathrm{tr}\left(\boldsymbol{\sigma}\right)\right)\mathrm{tr}\left(\dot{\boldsymbol{\sigma}}\right)=\mathrm{tr}\left(\dot{\boldsymbol{\sigma}}\right)/K \quad (29)$$

where $K=-\dfrac{1}{Jg'\left(\mathrm{tr}\left(\boldsymbol{\sigma}\right)\right)}$ is a function of the trace of the Cauchy stress $\boldsymbol{\sigma}$ or the volume ratio $J$. Noticing that this condition is in the current configuration, and therefore the unloading stress free configuration is not needed. We refer to Eq. (29) as the current rate form of the volume conservation condition.

Then we discuss how to revise an existing elastoplastic theory. For an arbitrary elastoplastic theory, its constitutive relationship can be expressed by

$$\dot{\boldsymbol{\sigma}}^{\mathrm{obj}}=\tilde{\boldsymbol{L}}:\boldsymbol{d} \quad (30)$$

where $\tilde{\boldsymbol{L}}$ is the tangential stiffness tensor and $\dot{\boldsymbol{\sigma}}^{\mathrm{obj}}$ represents one type of objective rate of the Cauchy stress $\boldsymbol{\sigma}$. If originally the constitutive relationship is expressed in other stress/strain rates than the Cauchy stress rate $\dot{\boldsymbol{\sigma}}^{\mathrm{obj}}$ and the deformation rate $\boldsymbol{d}$, it can be easily transformed to the form of Eq. (30), because the stress and the strain of different measures are related. As discussed previously, the elastoplastic theory is usually not volume conserved, so the volume conservation condition Eq. (24) or its rate form Eq. (29) is not satisfied with $\tilde{\boldsymbol{L}}$ as the tangential stiffness tensor. To realize volume conservation, we revise $\tilde{\boldsymbol{L}}$ to $\boldsymbol{L}^{rev}=\tilde{\boldsymbol{L}}+\boldsymbol{L}^{\delta}$ so the constitutive relationship Eq. (30) becomes

$$\dot{\boldsymbol{\sigma}}^{\mathrm{obj}}=\left(\tilde{\boldsymbol{L}}+\boldsymbol{L}^{\delta}\right):\boldsymbol{d} \quad (31)$$

Two conditions should be met in this revision: (1) after revision, Eq. (24) or its rate form Eq. (29) is satisfied; (2) the revision term $\boldsymbol{L}^{\delta}$ is the most minor one in the sense that its norm $\left\| \boldsymbol{L}^{\delta} \right\| = \sqrt{\boldsymbol{L}^{\delta} \vdots \boldsymbol{L}^{\delta}} = L_{ijkl}^{\delta} L_{ijkl}^{\delta}$ is the minimum.

The revision term of the stiffness tensor $\boldsymbol{L}^{\delta}$ depends on $\tilde{\boldsymbol{L}}$ and the choice on the objective stress rate. There are many different choices on the objective stress rate. Here we first use the Jaumann rate $\dot{\boldsymbol{\sigma}}^{Jau} = \dot{\boldsymbol{\sigma}} + \boldsymbol{\sigma} \cdot \boldsymbol{w} - \boldsymbol{w} \cdot \boldsymbol{\sigma}$ as an example, where $\boldsymbol{w}$ is the spin tensor. As demonstrated in Appendix F $\boldsymbol{L}^{\delta}$ is

$$\begin{cases} L_{iiii}^{\delta} = \dfrac{5}{9}\left(K - \sum_{k=1}^{3} \tilde{L}_{kkii}\right) - \dfrac{1}{9}\left(K - \sum_{k=1}^{3} \tilde{L}_{kkpp}\right) - \dfrac{1}{9}\left(K - \sum_{k=1}^{3} \tilde{L}_{kkqq}\right) & i = 1,2,3; p \neq i, q \neq i, p \neq q \\ L_{iijj}^{\delta} = L_{jjii}^{\delta} = \dfrac{2}{9}\left(K - \sum_{k=1}^{3} \tilde{L}_{kkii}\right) + \dfrac{2}{9}\left(K - \sum_{k=1}^{3} \tilde{L}_{kkjj}\right) - \dfrac{1}{9}\left(K - \sum_{k=1}^{3} \tilde{L}_{kkll}\right) & i,j = 1,2,3; i \neq j, l \neq i, l \neq j \\ L_{iikl}^{\delta} = L_{klii}^{\delta} = L_{iilk}^{\delta} = L_{lkii}^{\delta} = -\dfrac{1}{6}\left(\sum_{p=1}^{3} \tilde{L}_{ppkl} + \sum_{p=1}^{3} \tilde{L}_{pplk}\right) & i,k,l = 1,2,3; k \neq l \\ L_{ijkl}^{\delta} = L_{jikl}^{\delta} = L_{ijlk}^{\delta} = L_{klij}^{\delta} = 0 & i,j,k,l = 1,2,3; i \neq j, k \neq l \end{cases} \tag{32}$$

There is no dummy summation in Eq. (32). If an objective stress rate other than the Jaumann rate is used, the revision term of the stiffness tensor $\boldsymbol{L}^{\delta}$ can also be derived easily as presented in Appendix F.

In finite element simulations, we suggest the following procedure to carry out the revision and eliminate the possible accumulated numerical error that arises from the incremental algorithm.

Step 1: Supposing that at the current time, we have balanced stresses and compatible strains, then keep the strain unchanged, revise the stress according to Eq. (24) so that the deviation from the volume conservation condition is eliminated.

Step 2: the unbalanced stress caused by Step 1 is added to the total unbalanced force and finite element simulation is conducted using the incremental algorithm with the revised tangential stiffness tensor $\boldsymbol{L}^{rev}$ to obtain new balanced stresses and compatible strains. Then check if Eq. (24) is satisfied with an acceptable error; if not, return to Step 1.

We should notice that the choice of the objective stress rate is also an important issue in constitutive theories. Some of the objective stress rates, including the Jaumann rate, are not work-conjugate to a strain tensor (Bažant 1971, Ji et al. 2013), causing energy conservation problems. Due to this problem, an improper choice of the objective stress rates may lead to large errors when the material is highly compressible or highly anisotropic (Bažant et al., 2012; Ji et al., 2013; Vorel et al., 2013; Bažant and Vorel, 2014; Vorel and Bažant, 2014). It is recommended by Bažant that the Truesdell objective stress rate should be used instead of the commonly used Jaumann rate or the Green–Naghdi rate (Bažant and Vorel 2014,). However for materials such as metals, the error caused by using the Jaumann rate may still be ignored. A discussion of the choice of the objective stress rate in constitutive theory is not in the scope of the current paper and can be found in above literatures as well as in Ref. (Lee et al., 1983; Atluri, 1984; Xiao et al., 1997a, 1997b, 1998). We only emphasize that even though the elastoplastic theory chooses a proper objective rate, it still can have the volume conservation problem. Fortunately for an elastoplastic theory using an arbitrary objective stress, including the work-conjugate objective stress rate, the revision strategy presented in this subsection can still be carried out, as demonstrated in Appendix F.

### 4.3. New strategy 2, using a strain rate decomposition consistent with the volume conservation condition to establish a volume conserved theory

Through the procedure presented in Section 4.2, an arbitrary elastoplastic theory (including existing theories) can be revised so that the volume conservation is realized. After revision the plastic strain rate is usually no longer a deviatoric tensor. Because most elastoplastic theories are based on the assumption that the plastic strain rate is a deviatoric tensor, it will be convenient and consistent if the volume conservation condition can be satisfied while keeping the plastic strain rate a deviatoric tensor. In this subsection, a new strategy is proposed that: (1) does not need the unloading stress free configuration; (2) automatically satisfies the volume conservation without posterior revision and (3) keeps the plastic strain rate a deviatoric tensor.

To develop a strategy that meets the above requirements, we first introduce a strain rate decomposition consistent with the current deformation form of the volume conservation condition Eq. (24). An unloading configuration is usually necessary to realize the strain rate decomposition, and is preferred to be near the current configuration so that the unloading process is elastic (without reversal plastic deformation). The unloading path and the unloading configuration should be chosen so that the volume conservation condition Eq. (24) or its rate form Eq. (29) is satisfied, namely when the Cauchy stresses of two configurations are the same, their densities or volumes should be the same. In some theories this requirement is violated by choosing the unloading path and the unloading configuration in an arbitrary way, such as the Rice–Hill theories we discuss in Section 3. Therefore these theories cannot realize volume conservation. Here we show a way to choose the unloading configuration that is consistent with the volume conservation condition Eq. (24) or its rate form Eq. (29), as discussed below.

As the first step, a strain/stress measure needs to be chosen. According to Eq. (24) or Eq. (29), the most convenient stress measure seems to be the Cauchy stress because the density of volume is assumed to be a function of the trace of Cauchy stress. However, the Cauchy stress is defined in the current configuration so an objective stress rate also needs to be chosen among various candidates. To avoid this issue, we use the logarithmic stress $\boldsymbol{\sigma}^{\ln}$ and its work conjugate strain $\boldsymbol{E}^{\ln}$ instead, because the traces of the logarithmic stress $\boldsymbol{\sigma}^{\ln}$ and the Cauchy stress $\boldsymbol{\sigma}$ are closely related by $\mathrm{tr}\left(\boldsymbol{\sigma}^{\ln}\right)=J\mathrm{tr}\left(\boldsymbol{\sigma}\right)$. Besides the logarithmic stress is defined by taking the initial stress free configuration as the reference configuration so that the objective stress rate is avoided.

As illustrated in Fig. 10, after loading from configuration $\alpha$ to configuration $\beta$, the material volume element is unloaded to configuration $\chi$. Configuration $\chi$ is chosen so that it has the same logarithmic stress as configuration $\alpha$, i.e. $\boldsymbol{\sigma}^{\ln}_{(\chi)}=\boldsymbol{\sigma}^{\ln}_{(\alpha)}$. Then from Eq. (24) and $\mathrm{tr}\left(\boldsymbol{\sigma}^{\ln}\right)=J\mathrm{tr}\left(\boldsymbol{\sigma}\right)$, it can be proved that $\mathrm{tr}\left(\boldsymbol{\sigma}_{(\chi)}\right)=\mathrm{tr}\left(\boldsymbol{\sigma}_{(\alpha)}\right)$ and $\rho_{(\chi)}=\rho_{(\alpha)}$. Therefore this strain rate decomposition is consistent with the current deformation form of the volume conservation condition Eq. (24), and is not the same as that in the Rice–Hill theory, in which the unloading path is usually assumed to be parallel to the initial loading path or other artificially chosen one. This strain rate decomposition has a more solid physical background than those discussed in Section 3 in the sense that it is consistent with the volume conservation condition.

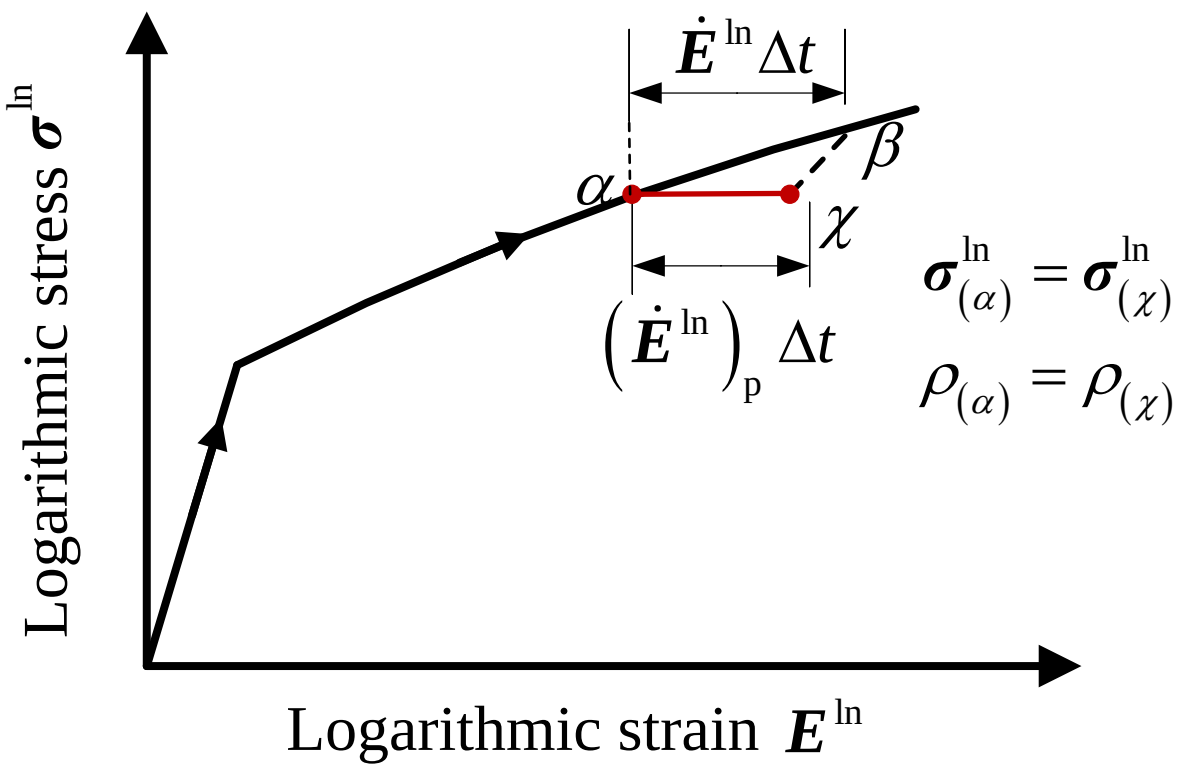

Fig. 10. The strain rate decomposition consistent with the volume conservation condition.

Both the Logarithmic stress and the density (or volume) of the configuration $\chi$ are equal to those of configuration $\alpha$.

Based on the idea discussed above, we establish an elastoplastic theory that can satisfy the volume conservation condition automatically. We take the combined hardening material as an example here, so the elastoplastic constitutive theory has the form

$$\left(\dot{\boldsymbol{E}}^{\ln}\right)_{\mathrm{p}} = \frac{9}{4C_{\mathrm{p}}\,\sigma_{\mathrm{eq}}^{2}}\left[\left(\boldsymbol{\sigma}^{\ln}\right)' - \left(\boldsymbol{\sigma}_{\mathrm{b}}\right)'\right] \otimes \left[\left(\boldsymbol{\sigma}^{\ln}\right)' - \left(\boldsymbol{\sigma}_{\mathrm{b}}\right)'\right] : \dot{\boldsymbol{\sigma}}^{\ln} \tag{33}$$

$$\left(\dot{\boldsymbol{E}}^{\ln}\right)_{\mathrm{e}} = \boldsymbol{M}_{\mathrm{e}} : \dot{\boldsymbol{\sigma}}^{\ln} \tag{34}$$

where $C_{\mathrm{p}}$ is the plastic modulus, $\boldsymbol{M}_{\mathrm{e}}$ is the elastic compliance tensor, $\boldsymbol{\sigma}_{\mathrm{b}}$ is the back stress, the superscript $'$ indicates the deviatoric part of a second order tensor, for example $\left(\boldsymbol{\sigma}^{\ln}\right)' = \boldsymbol{\sigma}^{\ln} - \frac{1}{3}\mathrm{tr}\left(\boldsymbol{\sigma}^{\ln}\right)\boldsymbol{I}$; $\sigma_{\mathrm{eq}} = \sqrt{\frac{3}{2}\left[\left(\boldsymbol{\sigma}^{\ln}\right)' - \left(\boldsymbol{\sigma}_{\mathrm{b}}\right)'\right] : \left[\left(\boldsymbol{\sigma}^{\ln}\right)' - \left(\boldsymbol{\sigma}_{\mathrm{b}}\right)'\right]}$; $\boldsymbol{I}$ is the second order identity tensor. From Eq. (33) we find that the plastic strain rate $\mathrm{tr}\left(\dot{\boldsymbol{E}}^{\ln}\right)_{\mathrm{p}} = 0$.

To realize the volume conservation by satisfying the volume conservation condition Eq. (24) or its rate form Eq. (29), the elastic compliance tensor $\boldsymbol{M}_{\mathrm{e}}$ should be determined correctly. Unlike in most other theories, in our theory the relation of the volume change vs. trace of Cauchy stress or the volume constitutive function $g\left(\mathrm{tr}\left(\boldsymbol{\sigma}\right)\right)$ in Eq. (24) plays an important role in determining $\boldsymbol{M}_{\mathrm{e}}$. In general $g\left(\mathrm{tr}\left(\boldsymbol{\sigma}\right)\right)$ in Eq. (24) should be determined from the experiment. Here we first present a linear form which is concise and reasonable in many cases. Realizing that for many materials like metals, the volume deformation is still modest even though the plastic deformation is large, it is reasonable to assume a linear relation of the volume change vs. the trace of Cauchy stress, namely

$$\frac{\Delta\rho}{\rho_{(0)}} = \frac{1}{J} - 1 = g\left(\mathrm{tr}\left(\boldsymbol{\sigma}\right)\right) = -\mathrm{tr}\left(\boldsymbol{\sigma}\right)/K_V \tag{35}$$

where $K_V$ is the constant volume modulus. To accurately satisfy the volume conservation condition Eq. (24) or Eq. (29), a stress dependent elastic compliance tensor $\boldsymbol{M}_{\mathrm{e}}$ is derived as

$$\boldsymbol{M}_{\mathrm{e}} = \frac{\left[K_V - \mathrm{tr}\left(\boldsymbol{\sigma}\right)\right]}{K_V^2\left(1-2\nu\right)}\left[\left(1+\nu\right)\mathscr{I} - \nu\boldsymbol{I}\otimes\boldsymbol{I}\right] \tag{36}$$

where $\mathscr{I}$ is the fourth order identity tensor. From Eq. (33) and Eq. (36), it is demonstrated that

$$\mathrm{tr}\left(\boldsymbol{d}\right) = \mathrm{tr}\left(\dot{\boldsymbol{E}}^{\ln}\right) = \frac{\left[K_V - \mathrm{tr}\left(\boldsymbol{\sigma}\right)\right]}{K_V^2}\mathrm{tr}\left(\dot{\boldsymbol{\sigma}}^{\ln}\right) = \frac{J}{K_V}\mathrm{tr}\left(\dot{\boldsymbol{\sigma}}\right) = -Jg'\left(\mathrm{tr}\left(\boldsymbol{\sigma}\right)\right)\mathrm{tr}\left(\dot{\boldsymbol{\sigma}}\right) \tag{37}$$

Therefore the volume conservation condition Eq. (29) is automatically satisfied. $\mathrm{tr}\left(\boldsymbol{\sigma}^{\ln}\right) = \mathrm{tr}\left(J\boldsymbol{\sigma}\right)$, $\mathrm{tr}\left(\boldsymbol{\sigma}^{\ln'}\boldsymbol{\sigma}^{\ln'} : \dot{\boldsymbol{\sigma}}^{\ln}\right) = 0$, Eq. (27) and Eq. (35) are used in above derivations. Finally the constitutive relationship is

$$\dot{\boldsymbol{E}}^{\ln} = \left\{\boldsymbol{M}_{\mathrm{e}} + \theta\frac{9}{4C_{\mathrm{p}}\,\sigma_{\mathrm{eq}}^{2}}\left[\left(\boldsymbol{\sigma}^{\ln}\right)' - \left(\boldsymbol{\sigma}_{\mathrm{b}}\right)'\right] \otimes \left[\left(\boldsymbol{\sigma}^{\ln}\right)' - \left(\boldsymbol{\sigma}_{\mathrm{b}}\right)'\right]\right\} : \dot{\boldsymbol{\sigma}}^{\ln} \tag{38}$$

where $\theta = 1$ during elastoplastic loading and $\theta = 0$ in the elastic range.

In this New Strategy 2, no information of the unloading process to zero stress is needed, so the elastoplastic theory avoids the unloading stress free configuration. The theory also avoids the choice of the objective stress rate, as the logarithmic strain and its work-conjugate stress is defined by taking the initial stress free configuration as the reference configuration.

From Eq. (35) and Eq. (36), we can see that when the relation of the volume change vs. trace of Cauchy stress is linear, the elastic compliance tensor $\boldsymbol{M}_e$ is not a constant. It is interesting to point out that the constant elastic compliance tensor $\boldsymbol{M}_e$ and the linear relation of the volume change vs. trace of Cauchy stress cannot be satisfied at the same time, because the relation between the volume and the strain is not linear for finite deformation.

The advantage of using the logarithmic strain and its work conjugate stress in elastoplastic theories has also been explored by others (Petric et al., 1992; Xiao et al., 1997a, 1997b; Bruhns et al., 1999; Xiao et al., 2000; Xiao et al., 2001; Arghavani et al., 2011). The theory of Petric et al. (1992) assumes a linear elastic compliance tensor and a relationship of the logarithmic stress/strain in the form of Eq. (38). It is a special case consistent with New Strategy 2, and is therefore volume conserved. Making use of the logarithmic strain rate and the Kirchhoff stress (Bruhns et al., 1999) or the logarithmic stress (Xiao et al., 2000), the Eulerian rate type elastoplastic theories are also in consistence with New Strategy 2 and therefore volume conserved. In many of the above theories, the constant elastic compliance tensor $\boldsymbol{M}_e$ is usually assumed. In our opinion, the linear relation of the volume change vs. trace of Cauchy stress is more reasonable and should have the priority to be satisfied. Some theories using logarithmic stress/strain also provide a general non-constant elastic compliance tensor $\boldsymbol{M}_e$, but it is still very difficult to satisfy the linear relation of the volume change vs. trace of Cauchy stress in their frames.

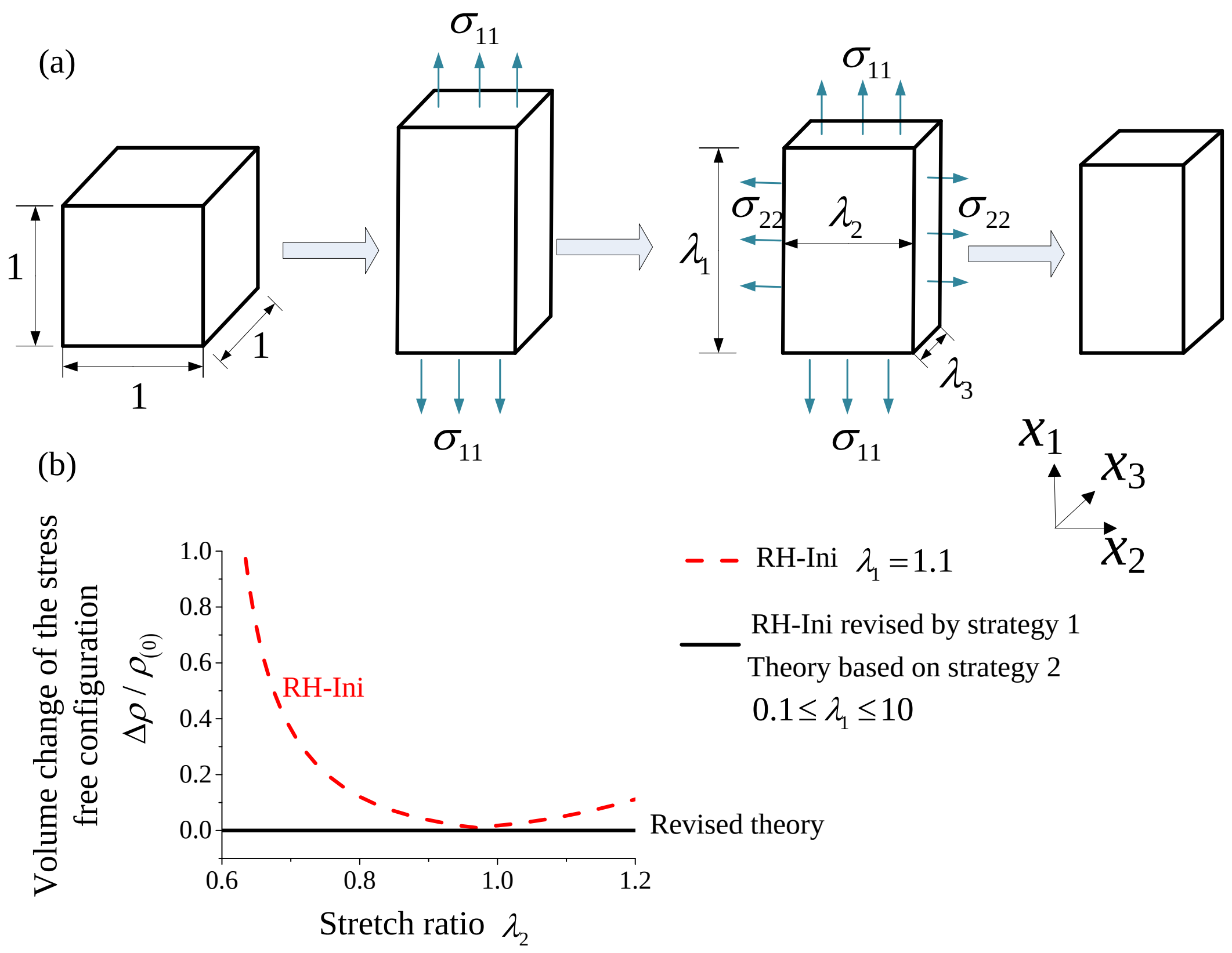

Fig. 11. The volume change of the stress free configuration of a material element subjected to subsequent biaxial loading, predicted by the linear hardening RH-Ini theories (the red dashed line) and the theories revised by New Strategy 1 and 2 (the black solid line).

### 4.4. Numerical examples to illustrate the volume conservation of New Strategy 1 and 2

based theory

The elastoplastic theory based on New Strategy 1 and 2 are evaluated by a numerical example where a material element is subjected to subsequent biaxial loadings ($\sigma_{11}$ is first applied and then fixed during $\sigma_{22}$ loading process) as shown in Fig. 11(a). The material property is still expressed by Eq.(20) and Eq.(21) (shown by Fig. 5).

The RH-ini theory (Rice–Hill strain rate decomposition with the initial configuration as the reference configuration) is revised by New Strategy 1 presented in Section 4.2. The predicted plastic volume deformation is almost zero, with a very small numerical error less than $10^{-8}$ for an increment $\lambda_{1t+\Delta t}/\lambda_{1t}=\mathrm{e}^{0.1}$ in each step, while the RH-ini theory without revision predicts a large plastic volume deformation even with a much smaller increment $\lambda_{1t+\Delta t}/\lambda_{1t}=\mathrm{e}^{0.001}$ as shown by the red dashed line in Fig. 11. The prediction using the elastoplastic theory based on New Strategy 2 also indicates almost zero plastic volume deformation, with a numerical error less than $10^{-10}$ for an increment $\lambda_{1t+\Delta t}/\lambda_{1t}=\mathrm{e}^{0.1}$.

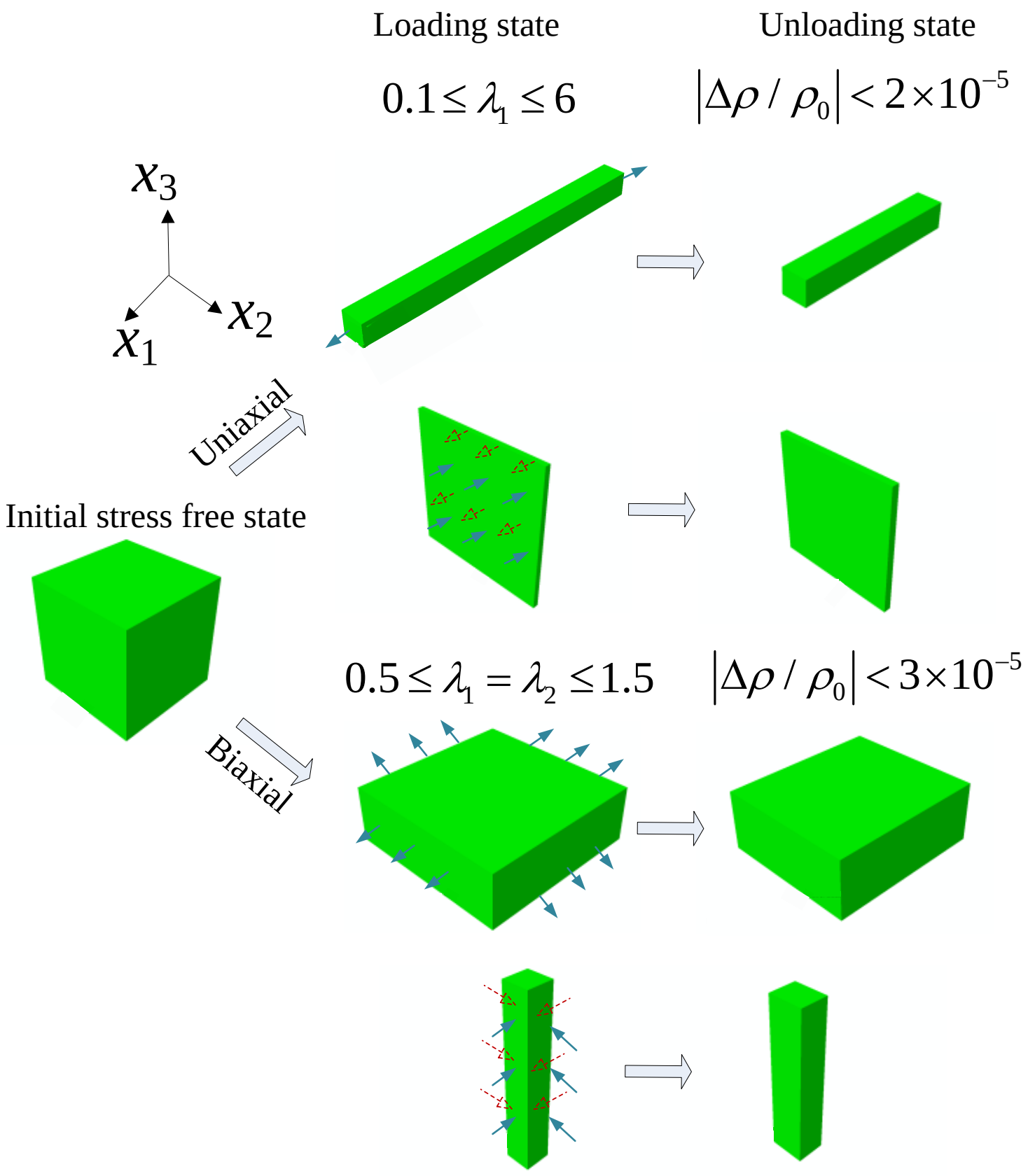

Fig. 12. Numerical examples of the New Strategy 2 based theory simulated by ABAQUS. The material property is linear combined hardening elastoplastic.

## 4.5. Numerical implement of the revised theory

To further illustrate the application of the revised theory based on the new strategies to realize volume conservation, the New Strategy 2 based elastoplastic theory is implemented into commercial software ABAQUS through the user material module. The material property is chosen to be combined hardening with the plastic modulus $C_{\mathrm{p}}=C_{\mathrm{e}}C_{\mathrm{ep}}/\left(C_{\mathrm{e}}-C_{\mathrm{ep}}\right)$ and the kinetic hardening modulus $C_{\mathrm{b}}=0.5C_{\mathrm{p}}$, so reverse plastic deformations occur in some of the following simulations. The numerical procedure to implement New Strategy 2 into ABAQUS is presented in Appendix G. Simulations on uniaxial loading and biaxial loading cases are carried out, and in all the cases the predicted volume deformations are negligible, as shown by Fig. 12.

## 5. Discussion

### 5.1. Applicability of the strategies to realize volume conservation

If the unloading stress free configuration can be uniquely determined from the current configuration, then by satisfying Eq. (14) the elastoplastic theory will be volume conserved. This strategy has been widely adopted in literature.

Two strategies to realize volume conservation without referring to the unloading stress free configuration are proposed in this paper. New Strategy 1 realizes the volume conservation by slightly revising the tangential stiffness tensor according to the relation between the density of material and the trace of the Cauchy stress tensor. This revision can be conducted on an arbitrary elastoplastic theory easily with just a little change to the original theory. After revision, the theory becomes strictly volume conserved for any stress free configurations, but might lead to slight incompatibility in some aspects, such as the plastic strain rate is not a deviatoric tensor anymore. However this cost is acceptable, as the volume conservation is fundamental in elastoplastic theories and some existing theories have serious problems in realizing volume conservation as demonstrated by the numerical examples in Section 3.3. The most advantageous aspect of this strategy is that it can be adopted for an arbitrary elastoplastic theory easily, no matter it depends on the unloading stress free configuration or not. An elastoplastic theory that is not volume conserved may have advantages in other aspects, such as a good prediction of the relationship between stresses and strains in certain cases. Therefore it may not be a good idea to abandon all the theories just for the volume conservation issue. New Strategy 1 can provide a remedy for these theories so that they can realize volume conservation.

A more rigorous treatment is provided by New Strategy 2. By the strain rate decomposition consistent with the current deformation form of the volume conservation condition, the logarithmic strain/stress measure and the properly determined elastic compliance tensor, the elastoplastic theory can then ensure the volume conservation condition while keeping the plastic strain rate still a deviatoric tensor. Neither the unloading stress free configuration nor the objective stress rate is needed in this strategy, because the logarithmic strain/stress used is defined by taking the initial stress free configuration as the reference configuration.

The comparison of the advantages and disadvantages among the three strategies are summarized in Table 2.

Table 2. Comparison among the strategies to realize volume conservation

| | Description | Advantage | Disadvantage |
|---|---|---|---|
| Strategy 0 | Satisfy the unloading stress free form of the volume conservation condition Eq. (14) by setting the proper plastic strain rate to be deviatoric | (1) without posterior revision;<br>(2) keeps the plastic strain rate a deviatoric tensor;<br>(3) directly applies for anisotropic media. | (1) needs the unloading stress free configuration;<br>(2) does not apply for materials with reversal plastic deformation. |
| New Strategy 1 | Slightly revise the stiffness tensor to satisfy the current form of the volume conservation condition Eq. (24) or Eq. (29) in the most minor way | (1) does not need the unloading stress free configuration;<br>(2) applies for materials with reversal plastic deformation;<br>(3) easy to implement for an arbitrary elastoplastic | (1) is posterior revision<br>(2) the plastic strain rate after revision may not be deviatoric;<br>(3) cannot directly apply for anisotropic media. |
| | | theory. | |
| New Strategy 2 | Use the strain rate decomposition consistent with the volume conservation condition, the logarithmic strain/stress and the properly determined elastic compliance tensor to satisfy Eq. (24) or Eq. (29) automatically | (1) does not need the unloading stress free configuration; (2) applies for materials with reversal plastic deformation; (3) without posterior revision; (4) keeps the plastic strain rate a deviatoric tensor. | (1) cannot directly apply for anisotropic media. |

5.2. Relation of the volume change vs. trace of Cauchy stress in Eq. (24)

In Section 4.3, a linear relation of the volume change vs. trace of Cauchy stress is assumed and the elastic compliance tensor is derived as in Eq. (36). It is reasonable for many materials such as metals, where the volume deformation is still modest even though the plastic deformation is large. Here we present a more general form for nonlinear relation of the volume change vs. trace of Cauchy stress. Supposing that we have already obtained the relation of the volume change vs. trace of Cauchy stress by experiment, the elastic compliance tensor $\boldsymbol{M}_{\mathrm{e}}$ is then derived as

$$\boldsymbol{M}_{\mathrm{e}} = -\frac{g'\left(\mathrm{tr}(\boldsymbol{\sigma})\right)\left[1+g\left(\mathrm{tr}(\boldsymbol{\sigma})\right)\right]}{\left[1+g\left(\mathrm{tr}(\boldsymbol{\sigma})\right)-g'\left(\mathrm{tr}(\boldsymbol{\sigma})\right)\mathrm{tr}(\boldsymbol{\sigma})\right](1-2\nu)}\left[(1+\nu)\mathscr{I}-\nu\boldsymbol{I}\otimes\boldsymbol{I}\right] \tag{39}$$

From Eq. (38) and Eq. (39), it is obtained that

$$\begin{aligned}\mathrm{tr}(\boldsymbol{d}) = \mathrm{tr}\left(\dot{\boldsymbol{E}}^{\ln}\right) &= -\frac{g'\left(\mathrm{tr}(\boldsymbol{\sigma})\right)\left[1+g\left(\mathrm{tr}(\boldsymbol{\sigma})\right)\right]}{\left[1+g\left(\mathrm{tr}(\boldsymbol{\sigma})\right)-g'\left(\mathrm{tr}(\boldsymbol{\sigma})\right)\mathrm{tr}(\boldsymbol{\sigma})\right]}\mathrm{tr}\left(\dot{\boldsymbol{\sigma}}^{\ln}\right) \\ &= \frac{g'\left(\mathrm{tr}(\boldsymbol{\sigma})\right)\left[1+g\left(\mathrm{tr}(\boldsymbol{\sigma})\right)\right]}{\left[1+g\left(\mathrm{tr}(\boldsymbol{\sigma})\right)-g'\left(\mathrm{tr}(\boldsymbol{\sigma})\right)\mathrm{tr}(\boldsymbol{\sigma})\right]}\left[Jg'\left(\mathrm{tr}(\boldsymbol{\sigma})\right)\mathrm{tr}(\boldsymbol{\sigma})-1\right]J\mathrm{tr}(\dot{\boldsymbol{\sigma}}) \\ &= -Jg'\left(\mathrm{tr}(\boldsymbol{\sigma})\right)\mathrm{tr}(\dot{\boldsymbol{\sigma}})\end{aligned} \tag{40}$$

Therefore the volume conservation condition Eq. (29) is automatically satisfied.

5.3. Small elastic deformation

Previous studies do not presume the degree of deformations. Considering that in some practical applications the elastic deformation is small, a brief discussion is presented below. If the elastic deformation gradient $\boldsymbol{F}^{\mathrm{e}}$ is close to the identity tensor, then from Eq. (16) and Eq. (18) it seems that SO-cur and MOS-sf theories are close to satisfy the volume conservation condition. However, we will demonstrate that the elastoplastic theories still cannot realize volume conservation within an acceptable error tolerance. We still investigate the example that a material element is subjected to uniaxial loading, but the material property is elastic-perfectly plastic now ($C_{\mathrm{ep}}=0$ in Eq. (20), Eq. (21) and Fig. 5). The maximum elastic strain (or yielding strain) here is about 0.1%. The volume changes of the unloading stress free configuration predicted by SO-cur theory, SO-ini theory, ABAQUS and the theories based on the new strategies proposed in this paper are presented in Fig. 13. We can see that with sufficient small increment, the SO-cur theory still predicts a plastic volume deformation of about 1.5% for $\lambda_1 = 0.1$ or $\lambda_1 = 10.0$, while ABAQUS predicts a plastic volume deformation of about 2.5% for $\lambda_1 = 0.1$ and 0.5% for $\lambda_1 = 10.0$, and the SO-ini theory has the worst prediction. Noticing that the error of the volume deformation is larger than the elastic deformation (on the order of 0.1%), which will lead to a significant error on the predicted stress. By contrast, the elastoplastic theory that implements the strategy proposed by this paper predicts almost zero plastic volume deformation (with a numerical error less than $10^{-5}$ for $0.1 \le \lambda_1 \le 10.0$ when implemented into ABAQUS).

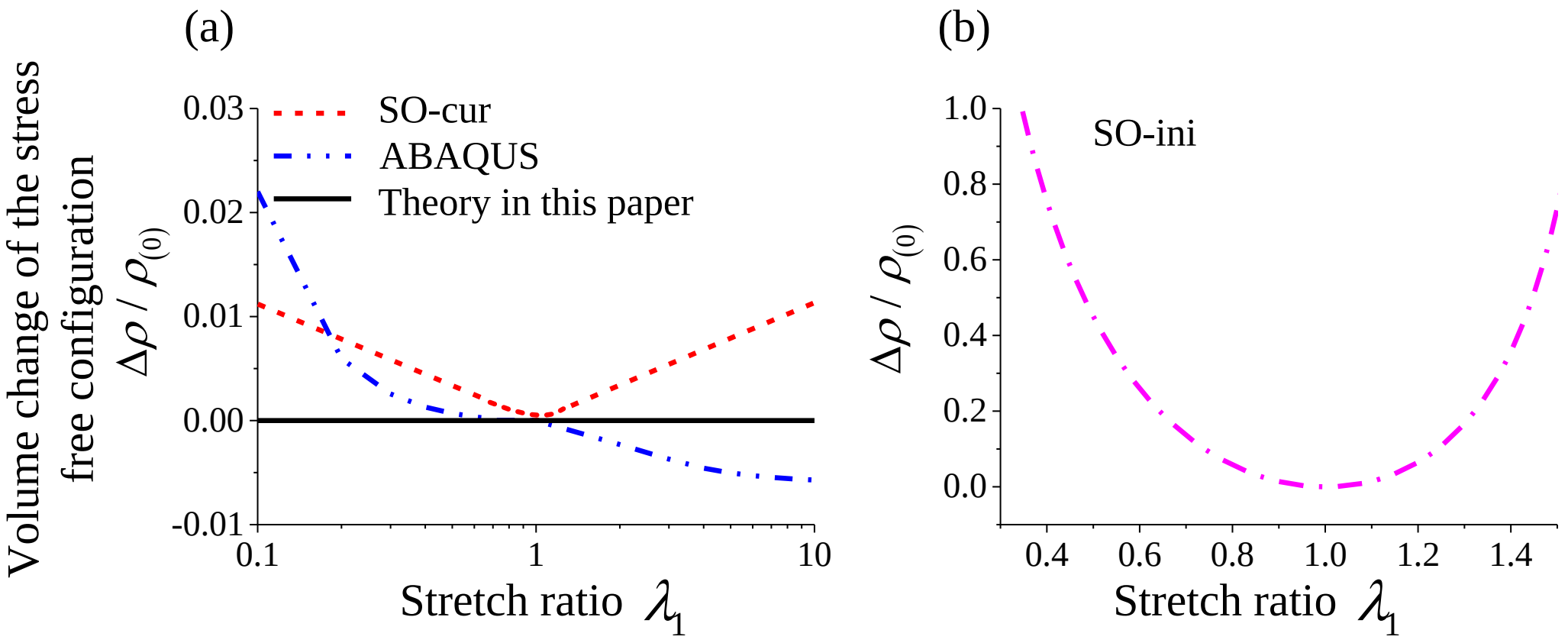

Fig. 13. The predicted volume change of the stress free configuration under uniaxial loading for elastic-perfectly plastic material with small elastic deformation.

### 5.4. Anisotropic media

For anisotropic media if no reversal plastic deformation occurs and the unloading stress free configuration can be uniquely determined, strategy 0 can still be used, i.e. satisfying the unloading stress free form of the volume conservation condition Eq. (14) by setting the proper plastic stress rate to be a deviatoric tensor. However the two New Strategies proposed in Section 4 are based on the assumption that the density or volume of material is a function of the trace of the Cauchy stress (Eq. (24)). While this assumption is reasonable for isotropic material, it needs to be revised for materials showing highly anisotropic behaviors, so are the formulae in the two New Strategies. This is a future direction of the current paper and will be presented in the authors' future paper. It may be interesting and important to revise the strategies for materials in various symmetry groups, such as transverse isotropy, orthotropy etc.

## 6. Conclusion

In this paper, we focus on the issue of rigorously realizing volume conservation during finite elastoplastic deformation. The following conclusions can be drawn.

(1) An unambiguous and accurate condition of volume conservation is clarified and used as benchmark, requiring that the density of any unloading states should be the same to that of the initial stress free state.

(2) Some classical elastoplastic theories are evaluated both theoretically and numerically. It is found that only two of them can realize volume conservation but they utilize the unloading stress free configuration and have several strong assumptions on the unloading behaviors. Except these two theories, numerical results indicate that the theories and software evaluated in this paper are not volume conserved and can predict very significant errors on plastic volume deformations.

(3) Based on a reasonable assumption that the density of material only depends on the trace of the Cauchy stress, two new strategies for volume conservation are thus proposed: directly and slightly revising the tangential stiffness tensor or using a properly chosen stress/strain measure and elastic compliance tensor. They are implemented in different elastoplastic theories, and the volume conservation is demonstrated by numerical examples. The established theories do not need the unloading stress free configuration and thus have expanded applicability. Their potential applications are an improved simulation of manufacture process such as metal forming.

(4) It is interestingly found that the zero trace of the plastic stress rate instead of the plastic strain rate is equivalent to the accurate condition of volume conservation. As this condition is very concise ( $\mathrm{tr}(\dot{\boldsymbol{\sigma}})_{\mathrm{p}}=0$ ), establishing a volume conserved elastoplastic theory based on the stress rate decomposition can be a promising future direction.

Acknowledgement

The authors acknowledge the support from National Natural Science Foundation of China (Grant Nos. 11425208, 51232004 and 11372158) and Tsinghua University Initiative Scientific Research Program (No. 2011Z02173).

Appendix A. The stress/strain measure

We adopt the Seth's strain measure here, which is written as

$$\boldsymbol{E}^{(\cdot)} = \sum_{i=1}^{3} \phi(\lambda_i) \boldsymbol{N}_i \boldsymbol{N}_i \tag{A1}$$

where $\boldsymbol{E}^{(\cdot)}$ is the strain using an arbitrary strain measure, $\boldsymbol{N}_i$ is the base vector of the Lagrange frame and $\phi$ is the measure function expressed by

$$\phi(\lambda) = \phi_n(\lambda) = \begin{cases} \dfrac{1}{2n}\left(\lambda^{2n} - 1\right) & n \neq 0 \\ \ln \lambda & n = 0 \end{cases} \tag{A2}$$

where $n$ is an arbitrary real number.

Appendix B. Evaluation of the volume conservation for more elastoplastic theories

As an extension of Section 3.2., this Appendix evaluates the volume conservation of more elastoplastic theories. Since the number of theories in literature is very large, we do not attempt to evaluate all of them. Instead we focus on some of the representative elastoplastic theories summarized and categorized in the review article by (Xiao et al., 2006).

*Classical Eulerian rate theories (In Section 5 of* Xiao et al., 2006*)*

Without referring to the unloading stress free configuration, the Eulerian rate theories decompose the deformation rate into $\boldsymbol{d} = \boldsymbol{d}_{\mathrm{e}} + \boldsymbol{d}_{\mathrm{p}}$. The plastic part $\boldsymbol{d}_{\mathrm{p}}$ is usually assume to be deviatoric, and the elastic part is related to the Kirchhoff stress $\boldsymbol{\tau}$ by

$$\boldsymbol{d}_{\mathrm{e}} = \frac{\partial^2 \Psi(\boldsymbol{\tau})}{\partial \boldsymbol{\tau} \partial \boldsymbol{\tau}} : \dot{\boldsymbol{\tau}}^{\mathrm{obj}} \tag{B1}$$

where $\Psi$ is the potential energy and $\dot{\boldsymbol{\tau}}^{\mathrm{obj}}$ denotes the objective rate of $\boldsymbol{\tau}$. Whether the theory is volume conserved or not depends on the choice of the potential energy $\Psi(\boldsymbol{\tau})$ and the objective stress rate. If the potential energy is expressed in the form $\Psi(\boldsymbol{\tau}) = \Psi\left(\mathrm{tr}(\boldsymbol{\tau}), \mathrm{tr}\left(\boldsymbol{\tau}^2\right)\right)$ and the objective stress rate is chosen to be the Jaumann rate, the Green–Naghdi rate or any rate that leads to $\mathrm{tr}\left(\dot{\boldsymbol{\tau}}^{\mathrm{obj}}\right) = \mathrm{tr}(\dot{\boldsymbol{\tau}})$, the elastoplastic theory is volume conserved because it satisfies the current rate form of the volume conservation condition Eq. (29) and is in consistent with New Strategy 2. However, if other objective rates such as the Truesdell rate, is used, the volume conservation cannot be guaranteed.

*Eulerian rate theory with the logarithmic rate (In Sec*tion 10 of Xiao et al., 2006)

A special case of the Eulerian rate theory is the one using the logarithmic rate as the objective stress rate. This theory is volume conserved as it is in consistence with New Strategy 2

*Lagrangean theory with plastic strain (In Section 6 of* Xiao et al., 2006*)*

The general form of this theory is expressed via the relationship between the Green strain $\boldsymbol{E}$ and the second Piola–Kirchhoff stress $\boldsymbol{S}$. The theory depends on the choice of three function, i.e. the yield function $Y\left(\boldsymbol{E}, \boldsymbol{E}^{\mathrm{p}}, \xi\right)$ in the strain space, the stress potential $\psi\left(\boldsymbol{E}, \boldsymbol{E}^{\mathrm{p}}, \xi\right)$ and the hardening function $\gamma\left(\boldsymbol{E}, \boldsymbol{E}^{\mathrm{p}}, \xi\right)$, where is $\boldsymbol{E}^{\mathrm{p}}$ the so-called plastic strain and $\xi$ is the internal variable. Due to its complicated form and the use of the Green strain instead of the logarithmic strain, the theory can hardly satisfy the volume conservation condition Eq. (24) or Eq. (29), and therefore is not volume conserved.

*Theories using the unloading stress free configuration (In Section 7 of Xiao et al., 2006)*

This class of theories differ from each other in the definition of the plastic strain rate via the plastic deformation gradient $\boldsymbol{F}^{\mathrm{p}}$, the elastic deformation gradient $\boldsymbol{F}^{\mathrm{e}}$ and their rates, as well as the choice of the reference configuration in which the plastic strain rate is assumed to be deviatoric. Besides the SO-sf and the MOS-cur theory discussed in Section 3 of this paper, the theories using the director triads and isoclinic configurations discussed in Section 7.6 of Xiao et al., 2006 is also volume conserved, as the flow rule is formulated for $\dot{\boldsymbol{F}}^{\mathrm{p}} \cdot (\boldsymbol{F}^{\mathrm{p}})^{-1}$ and can satisfy the unloading stress free form of the volume conservation condition Eq.(14) by setting $\dot{\boldsymbol{F}}^{\mathrm{p}} \cdot (\boldsymbol{F}^{\mathrm{p}})^{-1}$ to be deviatoric.

Appendix C. The Rice–Hill strain rate decomposition

Rice–Hill theory assumes that the plastic deformation origins from the internal structure evolvement of material. To introduce the Rice–Hill strain rate decomposition briefly, we consider a simple case that the internal structure can be described by one variable $\xi$ named as the internal variable, and other effects such as temperature change are ignored. The stress/strain measure is chosen to be the nominal strain and the nominal stress. Then the strain is a function of the stress and the internal variable, i.e. $\boldsymbol{E} = \boldsymbol{E}\left(\boldsymbol{\sigma}^{\mathrm{nominal}}, \xi\right)$. The plastic strain rate is defined as

$$\left(\dot{\boldsymbol{E}}^{\mathrm{RH}}\right)_{\mathrm{p}} = \frac{\partial \boldsymbol{E}\left(\boldsymbol{\sigma}^{\mathrm{nominal}}, \xi\right)}{\partial \xi} \dot{\xi} \tag{C1}$$

This definition is schematically illustrated by Fig. 2(a). The Rice–Hill plastic strain rate is not equal to the Simo–Ortiz plastic strain rate because the latter is

$$\left(\dot{\boldsymbol{E}}^{\mathrm{SO}}\right)_{\mathrm{p}} = \frac{\partial \boldsymbol{E}\left(0, \xi\right)}{\partial \xi} \dot{\xi} \tag{C2}$$

The difference between $\left(\dot{\boldsymbol{E}}^{\mathrm{RH}}\right)_{\mathrm{p}}$ and $\left(\dot{\boldsymbol{E}}^{\mathrm{SO}}\right)_{\mathrm{p}}$ is also illustrated by Fig. 2, noticing that the line AB and CD are usually not parallel. Therefore we have

$$\begin{cases} \left(\dot{\boldsymbol{E}}^{\mathrm{RH}}\right)_{\mathrm{p}} \neq \left(\dot{\boldsymbol{E}}^{\mathrm{SO}}\right)_{\mathrm{p}} \\ \left(\dot{\bar{\boldsymbol{E}}}^{\mathrm{RH}}\right)_{\mathrm{p}} = (\boldsymbol{F}^{\mathrm{p}})^{-\mathrm{T}} \cdot \left(\dot{\boldsymbol{E}}^{\mathrm{RH}}\right)_{\mathrm{p}} \cdot (\boldsymbol{F}^{\mathrm{p}})^{-1} \neq \left(\dot{\bar{\boldsymbol{E}}}^{\mathrm{SO}}\right)_{\mathrm{p}} \\ \left(\boldsymbol{d}^{\mathrm{RH}}\right)_{\mathrm{p}} = \boldsymbol{F}^{-\mathrm{T}} \cdot \left(\dot{\boldsymbol{E}}^{\mathrm{RH}}\right)_{\mathrm{p}} \cdot \boldsymbol{F}^{-1} \neq \left(\boldsymbol{d}^{\mathrm{SO}}\right)_{\mathrm{p}} \end{cases} \tag{C3}$$

Appendix D. Derivation of the plastic volume deformation of the uniaxial loading example

Considering the loading-unloading path from the initial stress free configuration through the current configuration and finally to the unloading stress free configuration, numerical calculations using increment algorithm are conducted to obtain the stress ratio $\lambda_{i(\mathrm{sf})}$ of the unloading stress free configuration and then the plastic volume deformation is obtained from its definition $\dfrac{\Delta\rho}{\rho_0} = \dfrac{1}{\lambda_{1(\mathrm{sf})}\lambda_{2(\mathrm{sf})}\lambda_{3(\mathrm{sf})}} - 1$.

D1. RH-ini theory

Supposing that in the current time point, the strain $\boldsymbol{E}^{\ln}$ and the stress $\boldsymbol{\sigma}$ are obtained, we derive the components of the strain rate $\dot{E}_{22}^{\ln} = \dot{E}_{33}^{\ln}$ as a function of the component $\dot{E}_{11}^{\ln}$. Then as we move forward along the loading–unloading path, the strain and the stress can be updated.

The reference configuration is chosen to be the initial configuration first. We consider the Rice–Hill strain rate decomposition using an arbitrary Seth's stress/strain measure $\boldsymbol{\sigma}^{(n)} / \boldsymbol{E}^{(n)}$ here and by taking $n=1$ it degenerates to the RH-ini theory of the main text. By taking $n=0$ it degenerates to the theory we discuss in Section 5.2. In the uniaxial loading example, the Seth's stress is related to the Cauchy stress by

$$\sigma_{11}^{(n)} = \lambda_2 \lambda_3 \frac{\sigma_{11}}{\lambda_1^{2n\text{-}1}} = \sigma_{11} \left\{ \exp\left[-(2n-1) E_{11}^{\ln}\right] + 2\exp\left(E_{22}^{\ln}\right) \right\} \tag{D1}$$

Now we derive the plastic and the elastic strain rate defined by the Rice–Hill theory during elastoplastic loading. The condition that the stress rate during elastoplastic loading equals to that of the elastic unloading leads to

$$\begin{aligned} &\left\{ \left[C_{\mathrm{ep}} - \sigma_{11}(2n-1)\right] \exp\left[-(2n-1)E_{11}^{\ln}\right] + 2C_{\mathrm{ep}} \exp\left(E_{22}^{\ln}\right) \right\} \dot{E}_{11}^{\ln} + 2\sigma_{11} \exp\left(E_{22}^{\ln}\right) \dot{E}_{22}^{\ln} \\ &= \left\{ \left[C_{\mathrm{e}} - \sigma_{11}(2n-1)\right] \exp\left[-(2n-1)E_{11}^{\ln}\right] + 2\left(C_{\mathrm{e}} - \nu\sigma_{11}\right) \exp\left(E_{22}^{\ln}\right) \right\} \left(\dot{E}_{11}^{\ln}\right)_{\mathrm{e}} \end{aligned} \tag{D2}$$

$\left(\dot{E}_{22}^{\ln}\right)_{\mathrm{e}} = -\nu \left(\dot{E}_{11}^{\ln}\right)_{\mathrm{e}}$ is used in the above derivation. The plastic strain rate $\boldsymbol{E}^{(n)}$ is a deviatoric tensor, so that

$$\left(\dot{E}_{22}^{(n)}\right)_{\mathrm{p}} = -\frac{1}{2} \left(\dot{E}_{11}^{(n)}\right)_{\mathrm{p}} \tag{D3}$$

Rewriting Eq.(D3) using the logarithmic strain and strain rate, we have

$$\left(\dot{E}_{22}^{\ln}\right)_{\mathrm{p}} = -\frac{1}{2} \exp\left(2nE_{11}^{\ln} - 2nE_{22}^{\ln}\right) \left(\dot{E}_{11}^{\ln}\right)_{\mathrm{p}} \tag{D4}$$

Equation (D2), Eq.(D4) and $\left(\dot{E}_{22}^{\ln}\right)_{\mathrm{e}} = -\nu \left(\dot{E}_{11}^{\ln}\right)_{\mathrm{e}}$ give rise to

$$\begin{aligned} \left(\dot{E}_{11}^{\ln}\right)_{\mathrm{e}} &= \left\{ \left[C_{\mathrm{e}} - \sigma_{11}(2n-1)\right] \exp\left[-(2n-1)E_{11}^{\ln}\right] - \sigma_{11} \exp\left[2nE_{11}^{\ln} - (2n-1)E_{22}^{\ln}\right] + 2C_{\mathrm{e}} \exp\left(E_{22}^{\ln}\right) \right\}^{-1} \\ &\times \left\{ \left[C_{\mathrm{ep}} - \sigma_{11}(2n-1)\right] \exp\left[-(2n-1)E_{11}^{\ln}\right] - \sigma_{11} \exp\left[2nE_{11}^{\ln} - (2n-1)E_{22}^{\ln}\right] + 2C_{\mathrm{ep}} \exp\left(E_{22}^{\ln}\right) \right\} \dot{E}_{11}^{\ln} \end{aligned} \tag{D5}$$

$$\begin{aligned} \left(\dot{E}_{11}^{\ln}\right)_{\mathrm{p}} &= \dot{E}_{11}^{\ln} - \left(\dot{E}_{11}^{\ln}\right)_{\mathrm{e}} \\ &= \left\{ \left[C_{\mathrm{e}} - \sigma_{11}(2n-1)\right] \exp\left[-(2n-1)E_{11}^{\ln}\right] - \sigma_{11} \exp\left[2nE_{11}^{\ln} - (2n-1)E_{22}^{\ln}\right] + 2C_{\mathrm{e}} \exp\left(E_{22}^{\ln}\right) \right\}^{-1} \\ &\times \left\{ \left(C_{\mathrm{e}} - C_{\mathrm{ep}}\right) \exp\left[-(2n-1)E_{11}^{\ln}\right] + 2\left(C_{\mathrm{e}} - C_{\mathrm{ep}}\right) \exp\left(E_{22}^{\ln}\right) \right\} \dot{E}_{11}^{\ln} \end{aligned} \tag{D6}$$

Therefore we have the desired relationship between the components of the strain rate, which are

$$\dot{E}_{22}^{\ln} = \dot{E}_{33}^{\ln} = -\frac{1}{2} \exp\left(2nE_{11}^{\ln} - 2nE_{22}^{\ln}\right) \left(\dot{E}_{11}^{\ln}\right)_{\mathrm{p}} - \nu \left(\dot{E}_{11}^{\ln}\right)_{\mathrm{e}} \tag{D7}$$

during elastoplastic loading and

$$\dot{E}_{22}^{\ln} = \dot{E}_{33}^{\ln} = -\nu \dot{E}_{11}^{\ln} \tag{D8}$$

during elastic loading or unloading.

Integrating Eq. (D7) and Eq. (D8) along the path indicated by the arrows in Fig. 5, we can obtain the logarithmic strain in the unloading stress free configuration.

$$E_{33(\mathrm{sf})}^{\ln} = E_{22(\mathrm{sf})}^{\ln} = \int \dot{E}_{22}^{\ln} \mathrm{d}\lambda_1 = \int \dot{E}_{22}^{\ln} \exp\left(E_{11}^{\ln}\right) \mathrm{d}E_{11}^{\ln} \tag{D9}$$

The stretch ratio is then calculated by $\lambda_{i(\mathrm{sf})} = \exp\left(E_{ii(\mathrm{sf})}^{\ln}\right)$, with no dummy summation on index *i*.

D2. RH-cur theory

When the current configuration is taken to be the reference configuration, the strain rate is the deformation rate $\boldsymbol{d}$ independent of the measure function, and the stress conjugate to $\boldsymbol{d}$ is the Kirchhoff stress $\boldsymbol{\tau} = J\boldsymbol{\sigma}$. Noticing that in uniaxial loading, the principal component of the deformation rate and the Kirchhoff stress is equal to the principal component of the logarithmic strain and the logarithmic stress respectively, i.e.

$$\tau_{ii} = \sigma_{ii}^{\ln}, d_{ii} = E_{ii}^{\ln} \quad \left(i = 1,2,3 \text{ no summation on } i\right) \tag{D10}$$

Therefore the derivation in Appendix D1 applies for the RH-cur theory by taking $n = 0$.

D3. Theories using Simo–Ortiz and Moran–Ortiz–Shih strain rate decompositions

Supposing that in the current time point, the strain $\boldsymbol{E}^{\ln}$ and the plastic deformation gradient $\boldsymbol{F}^{\mathrm{p}}$ are obtained, we derive the components of the rate of the plastic deformation rate $\dot{F}_{11}^{\mathrm{p}}$ and $\dot{F}_{22}^{\mathrm{p}} = \dot{F}_{33}^{\mathrm{p}}$ as a function of $\dot{E}_{11}^{\ln}$.

$F_{11}^{\mathrm{p}}$ is equal to the stretch ratio of the unloading stress free configuration

$$F_{11}^{\mathrm{p}} = \lambda_{1(\mathrm{sf})} = \exp\left(E_{11(\mathrm{sf})}^{\ln}\right) \tag{D11}$$

so

$$\dot{F}_{11}^{\mathrm{p}} = \dot{E}_{11(\mathrm{sf})}^{\ln} \exp\left(E_{11(\mathrm{sf})}^{\ln}\right) = \frac{C_{\mathrm{e}} - C_{\mathrm{ep}}}{C_{\mathrm{e}}} F_{11}^{\mathrm{p}} \dot{E}_{11}^{\ln} \tag{D12}$$

during elastoplastic loading and $\dot{F}_{11}^{\mathrm{p}} = 0$ during elastic loading or elastic unloading. The other two principal components of $\dot{\boldsymbol{F}}^{\mathrm{p}}$ are obtained from the zero trace of the plastic strain rates (Eq. (16) and (18)) as

$$\dot{F}_{22}^{\mathrm{p}} = \dot{F}_{33}^{\mathrm{p}} = \begin{cases} & \text{strain rate decomposition} & \text{refrence configuration} \\ -\dot{F}_{11}^{\mathrm{p}} F_{11}^{\mathrm{p}} / \left(2F_{22}^{\mathrm{p}}\right) & \text{Simo – Ortiz} & \text{Initial stress free} \\ -\dot{F}_{11}^{\mathrm{p}} F_{22}^{\mathrm{p}} / \left(2F_{11}^{\mathrm{p}}\right) & \text{Simo – Ortiz} & \text{Unloading stress free} \\ -\dot{F}_{11}^{\mathrm{p}} F_{22}^{\mathrm{p}} F_{22}^{\mathrm{e2}} / \left(2F_{11}^{\mathrm{p}} F_{11}^{\mathrm{e2}}\right) & \text{Simo – Ortiz} & \text{Current} \\ -\dot{F}_{11}^{\mathrm{p}} F_{11}^{\mathrm{p}} F_{11}^{\mathrm{e2}} / \left(2F_{22}^{\mathrm{p}} F_{22}^{\mathrm{e2}}\right) & \text{Moran – Ortiz – Shih} & \text{Initial stress free} \\ -\dot{F}_{11}^{\mathrm{p}} F_{22}^{\mathrm{p}} F_{11}^{\mathrm{e2}} / \left(2F_{11}^{\mathrm{p}} F_{22}^{\mathrm{e2}}\right) & \text{Moran – Ortiz – Shih} & \text{Unloading stress free} \\ -\dot{F}_{11}^{\mathrm{p}} F_{22}^{\mathrm{p}} / \left(2F_{11}^{\mathrm{p}}\right) & \text{Moran – Ortiz – Shih} & \text{Current} \end{cases} \tag{D13}$$

during elastoplastic loading and $\dot{F}_{22}^{\mathrm{p}} = \dot{F}_{33}^{\mathrm{p}} = 0$ during elastic loading or elastic unloading, where

$$F_{11} = \lambda_1 = \exp\left(E_{11}^{\ln}\right), F_{11}^{\mathrm{e}} = F_{11} / F_{11}^{\mathrm{p}} \tag{D14}$$

Integrating Eq.(D12) and Eq.(D13) along the path indicated by the arrows in Fig. 5, we can obtain $\boldsymbol{F}^{\mathrm{p}}$ in the unloading stress free configuration and the stretch ratios in the unloading stress free configuration are $\lambda_{i(\mathrm{sf})} = F_{ii}^{\mathrm{p}}$ where $i = 1,2,3$ with no dummy summation on the index.

Appendix E Numerical example of the incorrect stress predicted by some elastoplastic theories

If we attempt to shape a material element of unit size $1\times1\times1$ to a bar or plate whose width is $\lambda_{2(\mathrm{sf})}$, then from the stress–strain relation Eq. (20), Eq. (21) and the plastic volume conservation, we can derive that the applied stress in the current configuration should be

$$\sigma_{\text{accurate}} = \sigma_{\text{cr}} - 2\frac{C_{\text{e}}C_{\text{ep}}}{C_{\text{e}} - C_{\text{ep}}}\ln\left(\lambda_{2(\text{sf})}\right) \tag{E1}$$

when $\lambda_{2(\text{sf})} < 1$ (tension) and

$$\sigma_{\text{accurate}} = -\sigma_{\text{cr}} - 2\frac{C_{\text{e}}C_{\text{ep}}}{C_{\text{e}} - C_{\text{ep}}}\ln\left(\lambda_{2(\text{sf})}\right) \tag{E2}$$

when $\lambda_{2(\text{sf})} > 1$ (compression). But the elastoplastic theories and software that are not volume conserved predict a different applied stress, as shown by Fig. E1.

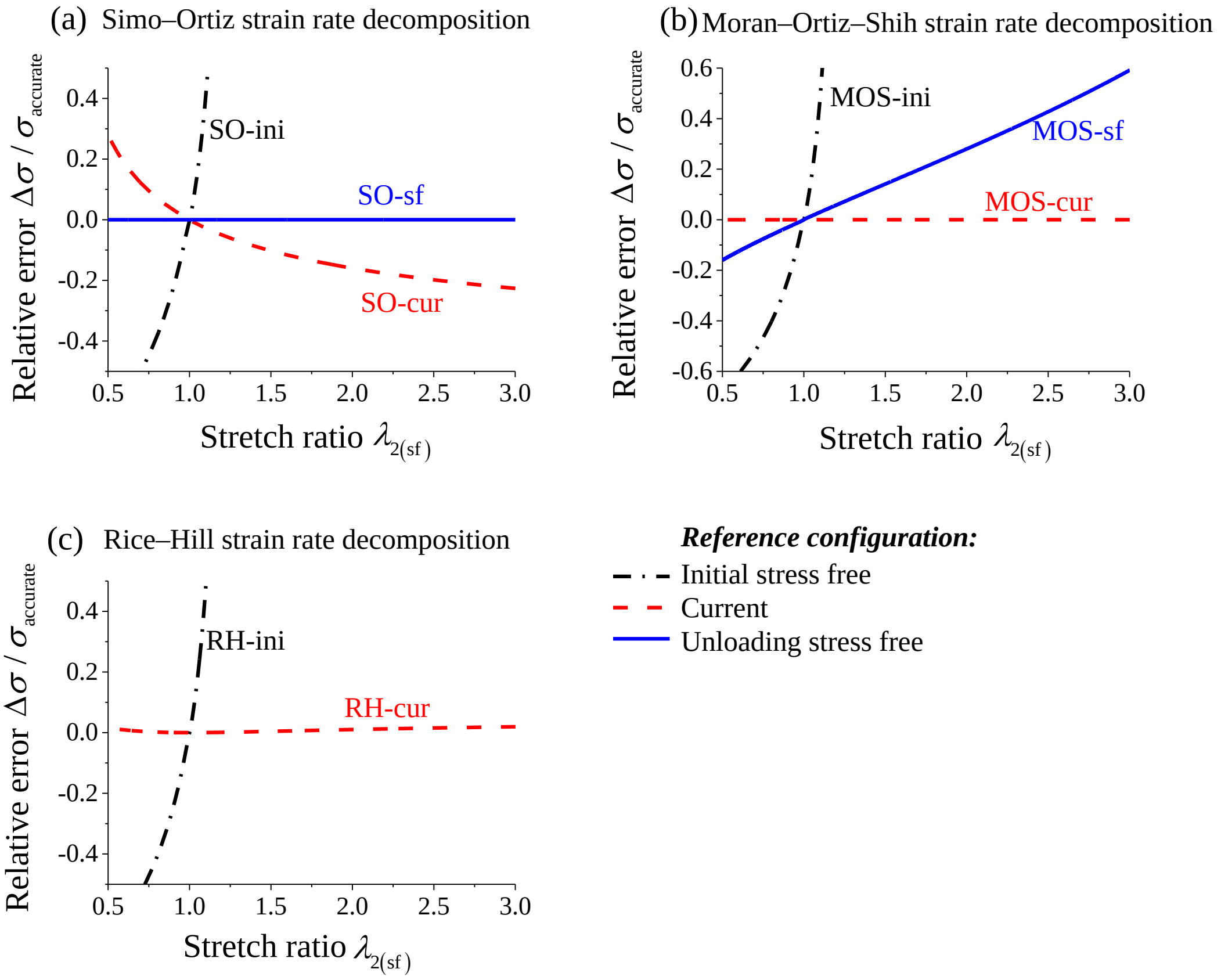

Fig. E1. The relative error of the predicted stress by some elastoplastic theories.

## Appendix F Derivation of the revision term in Eq. (31).

In this appendix we derive the revision term $\boldsymbol{L}^{\delta}$ to revise an elastoplastic theory which is originally not volume conserved. Adding the revision term to the original tangential stiffness tensor $\tilde{\boldsymbol{L}}$, we have

$$\dot{\boldsymbol{\sigma}}^{\text{obj}} = \left(\tilde{\boldsymbol{L}} + \boldsymbol{L}^{\delta}\right):\boldsymbol{d} \tag{F1}$$

There are various choices of the objective stress rates $\dot{\boldsymbol{\sigma}}^{\text{obj}}$, and here we derive $\boldsymbol{L}^{\delta}$ for some typical choices.

*Jaumann objective stress rate*

For the Jaumann rate $\dot{\boldsymbol{\sigma}}^{Jau} = \dot{\boldsymbol{\sigma}} + \boldsymbol{\sigma}\cdot\boldsymbol{w} - \boldsymbol{w}\cdot\boldsymbol{\sigma}$, we have

$$\sum_{i=1}^{3}\dot{\sigma}_{ii} = \sum_{k=1}^{3}\left[\left(\sum_{i=1}^{3}\tilde{L}_{iikk} + \sum_{i=1}^{3}L_{iikk}^{\delta}\right)d_{kk}\right] + \sum_{k,l=1,3;k\neq l}\left[\left(\sum_{i=1}^{3}\tilde{L}_{iikl} + \sum_{i=1}^{3}L_{iikl}^{\delta}\right)d_{kl}\right] \tag{F2}$$

Noticing that $\text{tr}\left(\boldsymbol{\sigma}\cdot\boldsymbol{w}\right) = 0$. The current rate form of the volume conservation condition requires that for any deformation rate $\boldsymbol{d}$ Eq.(29) is satisfied, so the revision term $\boldsymbol{L}^{\delta}$ is constrained by the following equations

$$\sum_{i=1}^{3}\tilde{L}_{ii11}+\sum_{i=1}^{3}L^{\delta}_{ii11}=\sum_{i=1}^{3}\tilde{L}_{ii22}+\sum_{i=1}^{3}L^{\delta}_{ii22}=\sum_{i=1}^{3}\tilde{L}_{ii33}+\sum_{i=1}^{3}L^{\delta}_{ii33}=K,\sum_{i=1}^{3}\tilde{L}_{iikl}+\sum_{i=1}^{3}L^{\delta}_{iikl}=0 \quad \text{(F3)}$$

$\boldsymbol{L}^{\delta}$ is also assumed to have the symmetric properties of an elastic stiffness tensor, namely $L^{\delta}_{ijkl}=L^{\delta}_{jikl}=L^{\delta}_{ijlk}=L^{\delta}_{klij}$. Then we seek for the most minor revision term $\boldsymbol{L}^{\delta}$ by minimizing its norm defined by $\left\|\boldsymbol{L}^{\delta}\right\|=\sqrt{\boldsymbol{L}^{\delta}\vdots\boldsymbol{L}^{\delta}}=\sum_{i,j,k,l=1,3}L^{\delta}_{ijkl}L^{\delta}_{ijkl}$, and the components presented in Eq.(32) are obtained.

*Work-conjugate objective stress rate*

A set of work-conjugate objective stree rate is proposed in (Bažant 1971) as

$$\overset{\circ}{\boldsymbol{\sigma}}^{(m)}=\dot{\boldsymbol{\sigma}}-(\boldsymbol{d}+\boldsymbol{w})\cdot\boldsymbol{\sigma}-\boldsymbol{\sigma}\cdot(\boldsymbol{d}-\boldsymbol{w})+\operatorname{tr}(\boldsymbol{d})\boldsymbol{\sigma}+\frac{1}{2}(2-m)(\boldsymbol{\sigma}\cdot\boldsymbol{d}+\boldsymbol{d}\cdot\boldsymbol{\sigma}) \quad \text{(F4)}$$

For $m=2$, Eq. (F4) reduces to the Truesdell objective stress rate. From Eq. (F1) and Eq. (F4) we have

$$\begin{aligned}\sum_{i=1}^{3}\dot{\sigma}_{ii}&=\sum_{k=1}^{3}\left[\left(\sum_{i=1}^{3}\tilde{L}_{iikk}-\sum_{i=1}^{3}\sigma_{ii}+m\sigma_{kk}+\sum_{i=1}^{3}L^{\delta}_{iikk}\right)d_{kk}\right]\\&+\sum_{k,l=1,3;k\neq l}\left[\left(\sum_{i=1}^{3}\tilde{L}_{iikl}+m\sigma_{kl}+\sum_{i=1}^{3}L^{\delta}_{iikl}\right)d_{kl}\right]\end{aligned} \quad \text{(F5)}$$

The same as in the case of Jaumann rate, for any deformation rate $\boldsymbol{d}$ Eq. (29) must be satisfied, $\boldsymbol{L}^{\delta}$ has the symmetry of an elastic stiffness tensor and is the most minor revision, so we derive that

$$\begin{cases}L^{\delta}_{iiii}=\dfrac{5}{9}\left(K-\sum_{k=1}^{3}\tilde{L}_{kkii}-m\sigma_{ii}\right)-\dfrac{1}{9}\left(K-\sum_{k=1}^{3}\tilde{L}_{kkpp}-m\sigma_{pp}\right)\\\quad-\dfrac{1}{9}\left(K-\sum_{k=1}^{3}\tilde{L}_{kkqq}-m\sigma_{qq}\right)+\dfrac{1}{3}\sum_{i=1}^{3}\sigma_{kk} & i=1,2,3;p\neq i,q\neq i,p\neq q\\L^{\delta}_{iijj}=L^{\delta}_{jjii}=\dfrac{2}{9}\left(K-\sum_{k=1}^{3}\tilde{L}_{kkii}-m\sigma_{ii}\right)+\dfrac{2}{9}\left(K-\sum_{k=1}^{3}\tilde{L}_{kkjj}-m\sigma_{jj}\right)\\\quad-\dfrac{1}{9}\left(K-\sum_{k=1}^{3}\tilde{L}_{kkll}-m\sigma_{ll}\right)+\dfrac{1}{3}\sum_{i=1}^{3}\sigma_{kk} & i,j=1,2,3;i\neq j,l\neq i,l\neq j\\L^{\delta}_{iikl}=L^{\delta}_{klii}=L^{\delta}_{iilk}=L^{\delta}_{lkii}=-\dfrac{1}{6}\left(\sum_{p=1}^{3}\tilde{L}_{ppkl}+\sum_{p=1}^{3}\tilde{L}_{pplk}+m\sigma_{kl}+m\sigma_{lk}\right) & i,k,l=1,2,3;k\neq l\\L^{\delta}_{ijkl}=L^{\delta}_{jikl}=L^{\delta}_{ijlk}=L^{\delta}_{klij}=0 & i,j,k,l=1,2,3;i\neq j,k\neq l\end{cases}$$

(F6)

Appendix G Numerical implement of the elastoplastic theory based on New Strategy 2

Step 0: before simulation, determine the initial yield stress, the plastic modulus $C_{\mathrm{p}}$, the kinetic hardening modulus $C_{\mathrm{b}}$, the isotropic hardening modulus $C_{\mathrm{F}}=C_{\mathrm{p}}-C_{\mathrm{b}}$ and $g(\operatorname{tr}(\boldsymbol{\sigma}))$ from uniaxial tests.

Step 1: calculate the strain increment $\Delta\boldsymbol{E}^{\ln}$. In ABAQUS the deformation gradient at the current time point $t$ and the trial deformation gradient at the subsequent time point $t+\Delta t$ are passed to the user subroutine (UMAT), denoted by $\boldsymbol{F}_{(t)}$ and $\boldsymbol{F}_{(t+\Delta t)}$ respectively here. The stretch ratio $\lambda_{i(t)}$, $\lambda_{i(t+\Delta t)}$ and the base vectors of Lagrange frame $\boldsymbol{N}_{i(t)}$, $\boldsymbol{N}_{i(t+\Delta t)}$ are obtained by calculating the eigenvalue and the principal direction of the right Cauchy-Green deformation tensor $\boldsymbol{C} = \boldsymbol{F}^{\mathrm{T}} \cdot \boldsymbol{F}$ . Then the strain increment is calculated as

$$\Delta \boldsymbol{E}^{\ln} = \sum_{i=1}^{3} \ln\left(\lambda_{i(t+\Delta t)}\right) \boldsymbol{N}_{i(t+\Delta t)} - \sum_{i=1}^{3} \ln\left(\lambda_{i(t)}\right) \boldsymbol{N}_{i(t)} \tag{G1}$$

Step 2: calculate the stress $\boldsymbol{\sigma}_{(t)}^{\ln}$ at the current time point. ABAQUS passes the Cauchy stress $\boldsymbol{\sigma}$ to the user subroutine, and it is converted to $\boldsymbol{\sigma}_{(t)}^{\ln}$.

Step 3: Check if the current increment is the elastic loading, elastoplastic loading or the elastic unloading. If the equivalent stress $\sigma_{\mathrm{eq}(t)}$ at time $t$ is less than the yielding stress $Y$, it is elastic loading or unloading; if $\sigma_{\mathrm{eq}(t)}$ equals to $Y$, calculate $\Delta \boldsymbol{\sigma}^{\ln} = \boldsymbol{M}_{\mathrm{e}}^{-1} : \Delta \boldsymbol{E}^{\ln}$, and if $\left[\left(\boldsymbol{\sigma}_{(t)}^{\ln}\right)' - \left(\boldsymbol{\sigma}_{\mathrm{b}(t)}^{\ln}\right)'\right] : \Delta \boldsymbol{\sigma}^{\ln} > 0$ it is elastoplastic loading, otherwise it is elastic unloading.

Step 4: Calculate the stress increment $\Delta \boldsymbol{\sigma}^{\ln}$. For elastic loading or elastic unloading $\Delta \boldsymbol{\sigma}^{\ln} = \boldsymbol{M}_{\mathrm{e}}^{-1} : \Delta \boldsymbol{E}^{\ln}$ ; for elastoplastic loading

$$\Delta \boldsymbol{\sigma}^{\ln} = \left\{ \boldsymbol{M}_{\mathrm{e}} + \frac{9}{4 C_{\mathrm{p}}\, \sigma_{\mathrm{eq}}^{2}} \left[\left(\boldsymbol{\sigma}^{\ln}\right)' - \left(\boldsymbol{\sigma}_{\mathrm{b}}\right)'\right] \otimes \left[\left(\boldsymbol{\sigma}^{\ln}\right)' - \left(\boldsymbol{\sigma}_{\mathrm{b}}\right)'\right] \right\}^{-1} : \Delta \boldsymbol{E}^{\ln} \tag{G2}$$

Step 5: Calculate the increment of the back stress.

Step 6: Calculate the Cauchy stress $\boldsymbol{\sigma}_{(t+\Delta t)}$ at time point $t + \Delta t$, from the logarithmic stress. Return the Cauchy stress back to ABAQUS, and it will decide whether the current state is in equilibrium. If not, ABAQUS will update the trial deformation gradient $\boldsymbol{F}_{(t+\Delta t)}$ so a new round of calculation starts from Step 1.